\documentclass[sigconf]{acmart}
\AtBeginDocument{\providecommand\BibTeX{{Bib\TeX}}}

\acmSubmissionID{38}

\usepackage{url}
\usepackage[ruled,vlined]{algorithm2e}
\usepackage{subfiles}
\usepackage{graphicx}
\usepackage{subcaption}
\usepackage{textcomp}
\usepackage[table]{xcolor}
\definecolor{stripe}{RGB}{254,254,217}
\usepackage{pifont}
\usepackage[utf8]{inputenc}
\usepackage{amsmath}
\usepackage{tikz}
\newcommand*\circled[1]{\tikz[baseline=(char.base)]{
            \node[shape=circle, draw, inner sep=1pt] (char) {#1};}}
\definecolor{exp-xMem}{HTML}{377eb8}
\definecolor{exp-DNNMem}{HTML}{e41a1c}
\definecolor{exp-SchedTune}{HTML}{4daf4a}
\definecolor{exp-LLMem}{HTML}{984ea3}
\definecolor{fq-bl}{HTML}{2C7BB6}
\definecolor{fq-br}{HTML}{FDAE61}
\definecolor{fq-tl}{HTML}{ABD9E9}
\definecolor{fq-tr}{HTML}{D7191C}

\def\BibTeX{{\rm B\kern-.05em{\sc i\kern-.025em b}\kern-.08em
    T\kern-.1667em\lower.7ex\hbox{E}\kern-.125emX}}
\usepackage{colortbl}
\usepackage{array}
\usepackage{booktabs}
\newcommand{\midsepremove}{\aboverulesep=0mm \belowrulesep=0mm}
\midsepremove
\newcommand{\midsepdefault}{\aboverulesep=0mm \belowrulesep=0mm}
\midsepdefault

\usepackage[inline]{enumitem}
	\setlist[enumerate]{leftmargin=1.2em}\setlist[itemize]{leftmargin=1.2em,label={$\bullet$}}

\newcommand{\projectname}{xMem\@\xspace}
\newcommand{\xmemtotalevaruns}{5209\@\xspace}

\newcommand{\xmemmedianerrorimprove}{91\%\@\xspace}
\newcommand{\xmemprobabilityimprove}{75\%\@\xspace}
\newcommand{\xmemmemoryimprove}{368\%\@\xspace}

\usepackage{xspace}

\newcommand*{\eg}{\textit{e.g.,}\@\xspace}

\makeatletter
\newcommand*{\etc}{\@ifnextchar{.}{etc}{etc.\@\xspace}}
\newcommand*{\etal}{\@ifnextchar{.}{et al}{et al.\@\xspace}}
\makeatother

\begin{document}

\acmYear{2025}\copyrightyear{2025}
\setcopyright{cc}
\setcctype[4.0]{by}
\acmConference[Middleware '25]{26th ACM Middleware Conference}{December 15--19, 2025}{Nashville, TN, USA}
\acmBooktitle{26th ACM Middleware Conference (Middleware '25), December 15--19, 2025, Nashville, TN, USA}
\acmDOI{10.1145/3721462.3770773}
\acmISBN{979-8-4007-1554-9/25/12}

\title[xMem: A CPU-Based Approach for Accurate GPU Memory Estimation in DL Training]{xMem: A CPU-Based Approach for Accurate Estimation of{\\}{ }GPU Memory in Deep Learning Training Workloads}

\author{Jiabo Shi}
\email{2865808S@student.gla.ac.uk}
\orcid{0009-0000-8326-3663}
\affiliation{\institution{University of Glasgow}
  \city{Glasgow}
  \country{UK}
}

\author{Dimitrios Pezaros}
\email{dimitrios.pezaros@glasgow.ac.uk}
\orcid{0000-0003-0939-378X}
\affiliation{\institution{University of Glasgow}
  \city{Glasgow}
  \country{UK}
}

\author{Yehia Elkhatib}
\email{yehia.elkhatib@glasgow.ac.uk}
\orcid{0000-0003-4639-436X}
\affiliation{\institution{University of Glasgow}
  \city{Glasgow}
  \country{UK}
}

\renewcommand{\shortauthors}{Shi, Pezaros, and Elkhatib}

\begin{abstract}
The global scarcity of GPUs necessitates more sophisticated strategies for Deep Learning jobs in shared cluster environments. 
Accurate estimation of how much GPU memory a job will require is fundamental to enabling advanced 
scheduling and GPU sharing, which helps prevent out-of-memory (OOM) errors and resource underutilization. However, existing estimation methods have limitations. Approaches relying on static analysis or historical data with machine learning often fail to accurately capture runtime dynamics. Furthermore, direct GPU analysis consumes scarce resources, and some techniques require intrusive code modifications. Thus, the key challenge lies in precisely estimating dynamic memory requirements, including memory allocator nuances, without consuming GPU resources and non-intrusive code changes. To address this challenge, we propose \projectname, a novel framework that leverages CPU-only dynamic analysis to accurately estimate peak GPU memory requirements a priori. We conducted a thorough evaluation of \projectname against state-of-the-art solutions using workloads from 25 different models, including architectures like Convolutional Neural Networks and Transformers. The analysis of \xmemtotalevaruns runs, which includes ANOVA and Monte Carlo results, highlights \projectname's benefits: it decreases the median relative error by \xmemmedianerrorimprove and significantly reduces the probability of estimation failure as safe OOM thresholds by \xmemprobabilityimprove, meaning that the estimated value can often be used directly without causing OOM. Ultimately, these improvements lead to a \xmemmemoryimprove increase in memory conservation potential over current solutions.

\end{abstract}

\ccsdesc[500]{Software and its engineering~Dynamic analysis}
\ccsdesc[300]{Computing methodologies~Machine learning}
\ccsdesc[500]{Software and its engineering~Extra-functional properties}

\keywords{
    GPU Memory Estimation,
    Dynamic Program Analysis,
    OOM Prevention,
    CPU-based Analysis,
    Resource Management, 
    GPU Cluster
}

\maketitle

\section{Introduction}
\label{introduction/chapter}
Deep Learning (DL) has experienced rapid growth, with Deep Neural Networks (DNNs) seeing widespread adoption across diverse domains (e.g. speech recognition~\cite{xiong_microsoft_2018}, image generation~\cite{Wang_2020_CVPR}, 
autonomous driving~\cite{grigorescu_survey_2020}, 
system capability description \cite{Zhang2023nlp}) 
and demonstrating performance comparable to human capabilities \cite{guerriero_iterative_2023}. As a result, technology companies are significantly increasing investments in high-performance computing infrastructure, particularly Graphics Processing Units (GPUs), which play a crucial role in accelerating model training. This surge in demand has led to an acute worldwide GPU shortage \cite{clark_great_2025}, which has affected the AI sector including prominent companies (such as OpenAI \cite{raza_openais_2023}), and contributed to technological and geo-economic competition tensions (\eg~\cite{anthropic2025framework}). 

To maximize the utilization of these scarce and expensive resources, DL training increasingly relies on shared GPU clusters \cite{276938}. However, managing shared environments introduces significant complexities in resource allocation and job scheduling \cite{222611, MLSYS2020_d9cd83bc, kokolis_revisiting_2025}. Among these, effective GPU memory management emerges as a particularly critical challenge. Inefficient allocation or unpredictable memory demands frequently lead to underutilization \cite{288717, gu_liquid_2022, li_miso_2022, 211297, 276938, li_lyra_2023, pavlidakis_guardian_2024, zhang_rubick_2024} and disruptive \emph{Out-of-Memory (OOM) errors}, identified as a major cause of job failures in production clusters \cite{cheng_towards_2023, zhang_empirical_2020}. Therefore, accurate GPU memory estimation is essential to mitigate these costly OOM failures, optimize scheduling, and improve overall cluster efficiency and stability. Specifically, precise GPU memory estimation for DL jobs enables decision systems within shared GPU clusters (of cloud providers, research institutions, and enterprises) to implement more effective and intelligent resource scheduling, leading to substantial GPU memory conservation and optimized asset utilization, which in turn helps mitigate the prevailing GPU scarcity.

\begin{figure}[tbp]
    \centering
    \includegraphics[width=\linewidth]{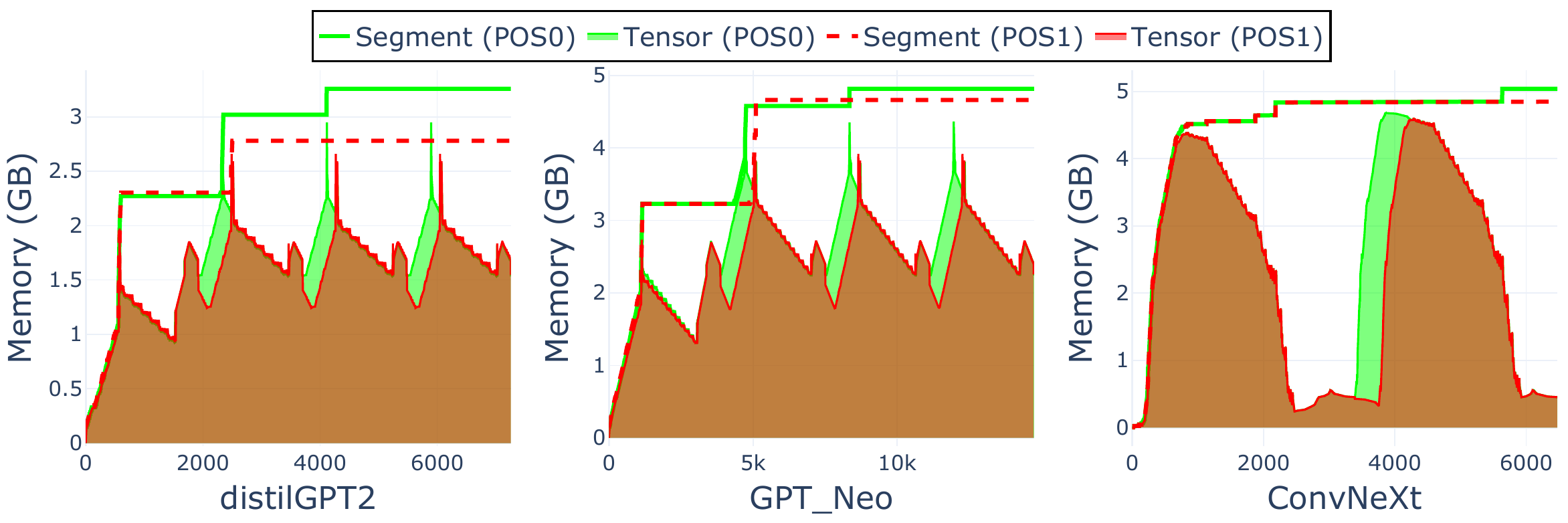}
    \caption{The impact of \texttt{optimizer.zero\_grad()} placement on GPU memory between calling \texttt{zero\_grad()} before backward propagation (POS0) versus at the start of iteration (POS1), illustrating runtime sensitivity and allocator impact. While the underlying Tensor memory activity (green/red areas for POS0/POS1) might show similar patterns, the overall Segment memory footprint (solid and dashed lines) differs significantly.}
    \Description{Changing the position of the code, such as \texttt{optimizer.zero\_grad()}, also impacts peak memory usage during a training job due to the nature of the dynamic runtime.}
    \label{motivation/fig/zero-out-memory-change}
\end{figure}

However, developing estimation methodologies that are both accurate and practical for ML training is challenging.\begin{enumerate*}[label=]\item \textbf{Firstly}, achieving high estimation accuracy is inherently difficult due to the complex and dynamic runtime behavior of GPU memory usage. Factors such as intermediate tensor lifetimes, dynamic operator execution, intricate workings of memory allocators, and even subtle code details such as the precise placement of gradient zeroing operations within a training loop (see Figure~\ref{motivation/fig/zero-out-memory-change}) can significantly alter memory consumption patterns, making them difficult to model precisely based on a static computation graph \cite{gao_estimating_2020}.\item \textbf{Secondly}, for these estimations to be useful in guiding group scheduling decisions, they must be available a priori, before the start of the job execution, imposing a strict constraint crucial to proactive scheduling and resource allocation \cite{li_miso_2022, yeung_horus_2022, li_lyra_2023}.\item \textbf{Thirdly}, given the scarcity of GPU resources, it is imperative that the estimation process itself operates without consuming target GPU resources, thereby avoiding additional overhead and contention. 
\end{enumerate*}
Finding an estimation method that can accurately handle runtime dynamics, provide results before the job starts, require no code changes, and avoid using the target GPU entirely remains the central challenge for reliable and practical GPU memory estimation.

Despite the clear need, existing methodologies for GPU memory estimation struggle to simultaneously satisfy the demanding requirements of high accuracy for dynamic workloads, a priori availability, and zero target-GPU overhead. Static analysis approaches, for instance, examine the computational graph of a model before execution, \eg DNNMem \cite{gao_estimating_2020} and Horus \cite{yeung_horus_2022}. However, they are fundamentally limited in their ability to capture crucial runtime dynamics. These methods often fail to accurately account for the complex behavior of memory allocators \cite{pytorch_cuda_2024} or the significant impact of subtle code changes, such as the placement of gradient zeroing operations (Figure~\ref{motivation/fig/zero-out-memory-change}), on memory sequences and overall consumption. Consequently, their accuracy can be compromised, particularly for complex models or during the iterative model development cycle, which involves frequent changes.

Alternatively, data-driven techniques leverage historical execution data or model features to train machine learning estimators, \eg SchedTune \cite{albahar_schedtune_2022}, TBEM \cite{liu_tbem_2022}. Although these can provide a priori estimations, they often face challenges with generalization to unseen models (cold start problem \cite{talha_deep_2023}) and may not precisely capture instantaneous peak memory demands. Another category involves direct GPU interaction through profiling, testing, or partial execution, \eg LLMem \cite{kim_llmem_2024}, Gandiva's online profiling \cite{222611}. While these approaches may potentially offer higher accuracy, they directly violate the critical zero target-GPU overhead constraint by consuming valuable GPU resources and adding significant overhead, rendering them unsuitable for fast, pre-scheduling decisions or environments without direct GPU access. Furthermore, some techniques in this category, reportedly including LLMem \cite{kim_llmem_2024}, may require intrusive modifications to the user's training code, hindering practical adoption due to increased developer effort and maintainability concerns.

Therefore, a significant gap remains: no existing type of solution consistently delivers high-accuracy, dynamic-aware memory predictions a priori without resorting to costly target GPU usage and non-intrusive code changes. This motivates the search for novel approaches. To bridge this gap, we propose \projectname, a novel framework designed to accurately estimate the peak GPU memory consumption for DL training jobs using data from a profiler \cite{pytorch_pytorch_2024}, a performance analyzer frequently used during DL model development, thereby eliminating the need for code changes.

\projectname fundamentally departs from GPU-dependent methods by employing entirely CPU-based dynamic analysis. The feasibility of this approach is underpinned by key observations regarding DL framework execution, specifically using PyTorch as our major example: \label{intro/paragraph/key-observations}
\begin{enumerate*}[label=(\roman*)]
    \item reflecting the design goal of deep learning frameworks to execute the same essential training task across different hardware. For example,  the high-level logic defined in the Python training script determines a consistent set and count of core tensors, \eg parameters, activations, gradients, required per iteration, irrespective of the backend (CPU vs. GPU). Although the specific memory footprint and precise lifecycle timing during operator processing can differ based on varying low-level implementations, such as using MKL vs. cuDNN, allocator behavior, or hardware characteristics \cite{abadi_tensorflow_2016, paszke_pytorch_2019};
    \item substantial optimization efforts have been invested in common operators for both CPU and GPU platforms, frequently resulting in their core memory footprints being closely comparable during execution; and
    \item the behavior of crucial components like memory allocators (e.g., PyTorch's CUDACachingAllocator \cite{pytorch_cuda_2024}) follows deterministic logic that can be simulated if the input sequence of memory events is known, even if derived from CPU execution.
\end{enumerate*}

Based on these principles, \projectname utilizes a CPU-based estimation framework that combines CPU trace analysis, memory orchestration, and two-level allocator simulation to estimate peak GPU memory usage before execution. We rigorously evaluated our approach using Analysis of Variance (ANOVA)~\cite{scheffe_analysis_1999} and Monte Carlo simulation \cite{rubinstein_simulation_2017}, conducting approximately \xmemtotalevaruns runs across 25 models. Our evaluation encompassed a range of model architectures including modern Convolutional Neural Networks (CNNs) like ConvNeXt \cite{liu_convnet_2022} and recent Transformer models such as Qwen3 \cite{yang_qwen3_2025}. Comparing against recent solutions --- DNNMem \cite{gao_estimating_2020}, SchedTune \cite{albahar_schedtune_2022}, and LLMem \cite{kim_llmem_2024} --- \projectname demonstrates significant improvements: it decreases Median Relative Error (MRE) by \xmemmedianerrorimprove, reduces Probability of Estimation Failure (PEF) by \xmemprobabilityimprove, and increases Memory Conservation Potential (MCP) by \xmemmemoryimprove. 
These results validate the effectiveness of our CPU-based methodology and its practical benefits. In this paper, we make the following key contributions: 
\begin{enumerate}
    \item We demonstrate the feasibility of accurately estimating peak GPU memory for DL training a priori using solely CPU-based dynamic analysis.
    \item We present \projectname, a novel framework that embodies this approach, detailing its multistage pipeline involving CPU trace analysis, memory orchestration, and allocator simulation.
    \item We empirically show that \projectname achieves superior performance when compared to three state-of-the-art methods: significantly increasing accuracy of predicting memory usage, reducing the risk of OOM, and improving memory conservation.
    \item  We release \projectname as open source
    \footnote{The \projectname source code can be publicly accessed via \href{https://github.com/Stone-ResearchLife/xMem}{this repository}.\label{introduction/foot/source-code}} 
    to the research community, including a valuable Python simulator for the CUDACachingAllocator, to promote adoption and facilitate future work.
\end{enumerate} 

\section{Background}
\label{background/chapter}
\subsection{Out-of-Memory Issues} \label{background/chapter/OOM-issue}

Empirical studies identify OOM as a major error category in DL development \cite{islam_comprehensive_2019}, especially worsening in resource-constrained GPU clusters. Nonetheless, even with datacenter-grade GPUs offering larger memory, production clusters at both Microsoft and Meta have attributed 9\% of DL training failures to OOM \cite{cheng_towards_2023, zhang_empirical_2020}. These challenges stem from the high memory demands of training DL models, exceeding the limited memory capacity available on GPUs. Furthermore, accurately managing or estimating GPU memory usage is inherently difficult because it is highly sensitive to various factors including training batch size, the choice of optimizer, and even minor code alterations made during the dynamic process of model development and tuning.

\subsection{Memory Consumption} \label{background/chapter/memory}
Understanding memory usage is crucial for the accurate estimation of the model training job. Typically, memory consumption within a training job can be categorized into two key types: memory used directly by Tensors and memory managed by the underlying allocator, namely Segments. We will be focusing on PyTorch as a primary example.

\subsubsection{Tensors} \label{background/chapter/tensor-memory}
Tensors represent the core data structures in PyTorch  \cite{paszke_pytorch_2019}, holding model parameters, activations, gradients, and other intermediate values necessary during training. The memory required for these tensors is dynamic and is influenced by factors such as the model architecture, batch size, and allocator memory alignment strategy.

\begin{figure}[htbp]
    \centering
    \includegraphics[width=\linewidth]{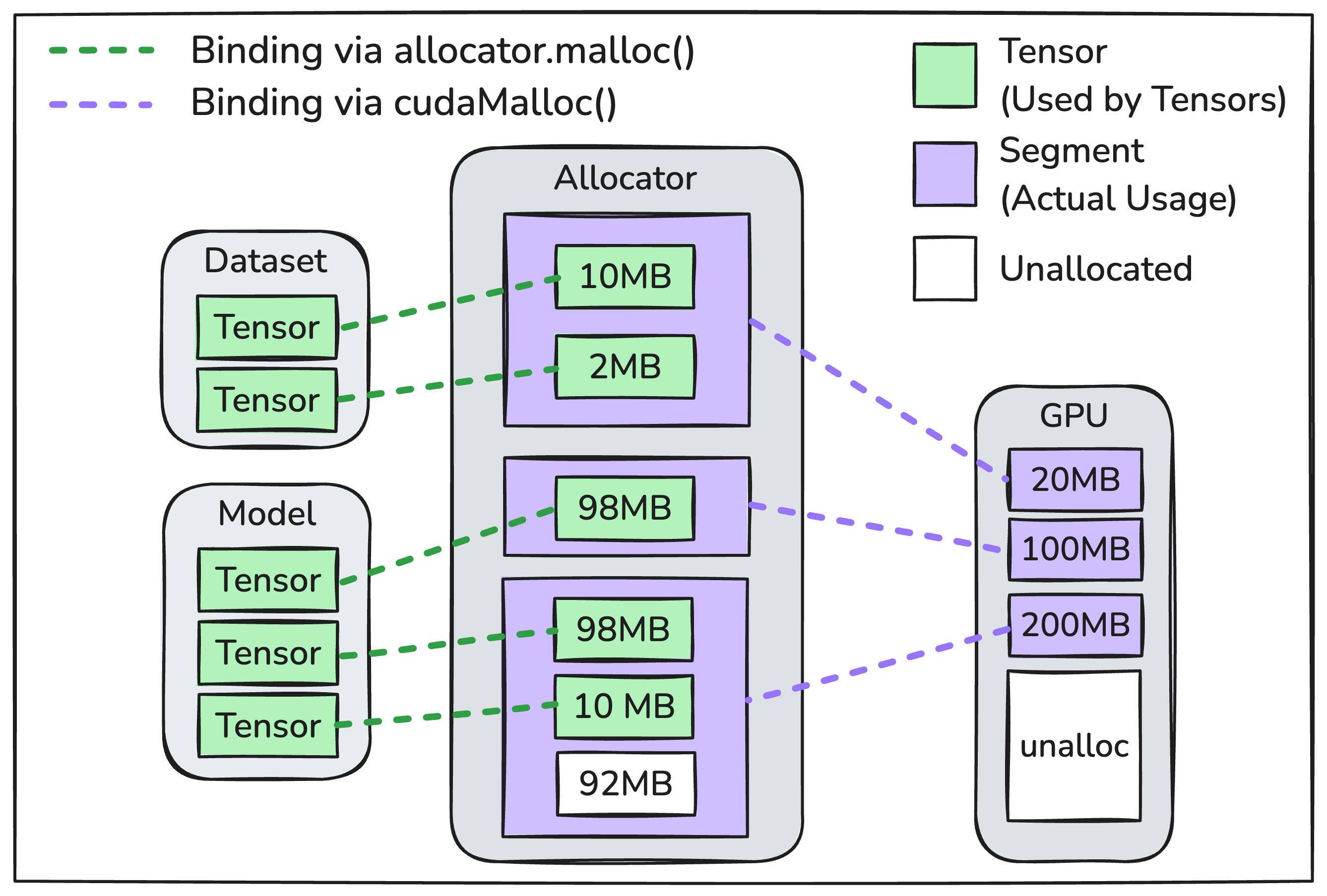}
    \caption{An example of how memory blocks are managed in PyTorch for a training job.}
    \Description{Diagram illustrates how tensors from datasets and models are bound to GPU memory segments via a framework using different allocation methods.}
    \label{background/fig/memory-management}
\end{figure}

It is important to note that tensor memory is not allocated directly from the GPU's raw memory pool. Instead, frameworks such as PyTorch employ a caching allocator (specifically, CUDACachingAllocator \cite{pytorch_cuda_2024} is the default for CUDA devices), which preallocates larger blocks of GPU memory, known as Segments, described next.
This allocator then serves tensor memory requests by splitting blocks from these preallocated Segments. As illustrated in Fig.~\ref{background/fig/memory-management}, the memory blocks allocated for individual tensors (represented by solid green rectangles) reside within larger Segments (purple boxes). Furthermore, to ensure efficient memory access and meet hardware alignment requirements, PyTorch's allocator often rounds up the requested memory size for a tensor; \eg 
to the next multiple of 512 bytes.

\subsubsection{Segments} \label{background/chapter/segment}
Segments are larger blocks of memory requested directly from the GPU by PyTorch's caching allocator. These act as reservoirs from which memory is allocated for individual tensors. When a tensor is no longer needed, its memory block is marked as deallocated within its segment but is not immediately returned to the GPU. The allocator retains these deallocated blocks (cached memory) for potential reuse by future tensor allocations of compatible sizes. This caching strategy significantly reduces the communication overhead with the GPU driver.

The allocator requests a new Segment from the GPU only when the existing segments are insufficient. Again to minimize GPU communication overhead, the allocator often requests more memory than immediately needed, \eg, requesting 2MB for a 1MB tensor need, and caches the excess. The size of these requested Segments is determined by the allocator's internal strategy balancing speed and fragmentation, as denoted by the purple boxes in Fig.~\ref{background/fig/memory-management}. Consequently, the total GPU memory consumption by an estimation should primarily reflect the sum of the sizes of all segments allocated from the GPU, not just the memory actively used by tensors at any given moment. Overlooking the behavior of this caching allocator, as in some prior studies \cite{yeung_horus_2022, albahar_schedtune_2022}, can lead to inaccurate estimations of peak GPU memory requirements during training.

\subsection{Sequence of Memory Activities}

\begin{figure}[htbp]
    \centering
    \includegraphics[width=\linewidth]{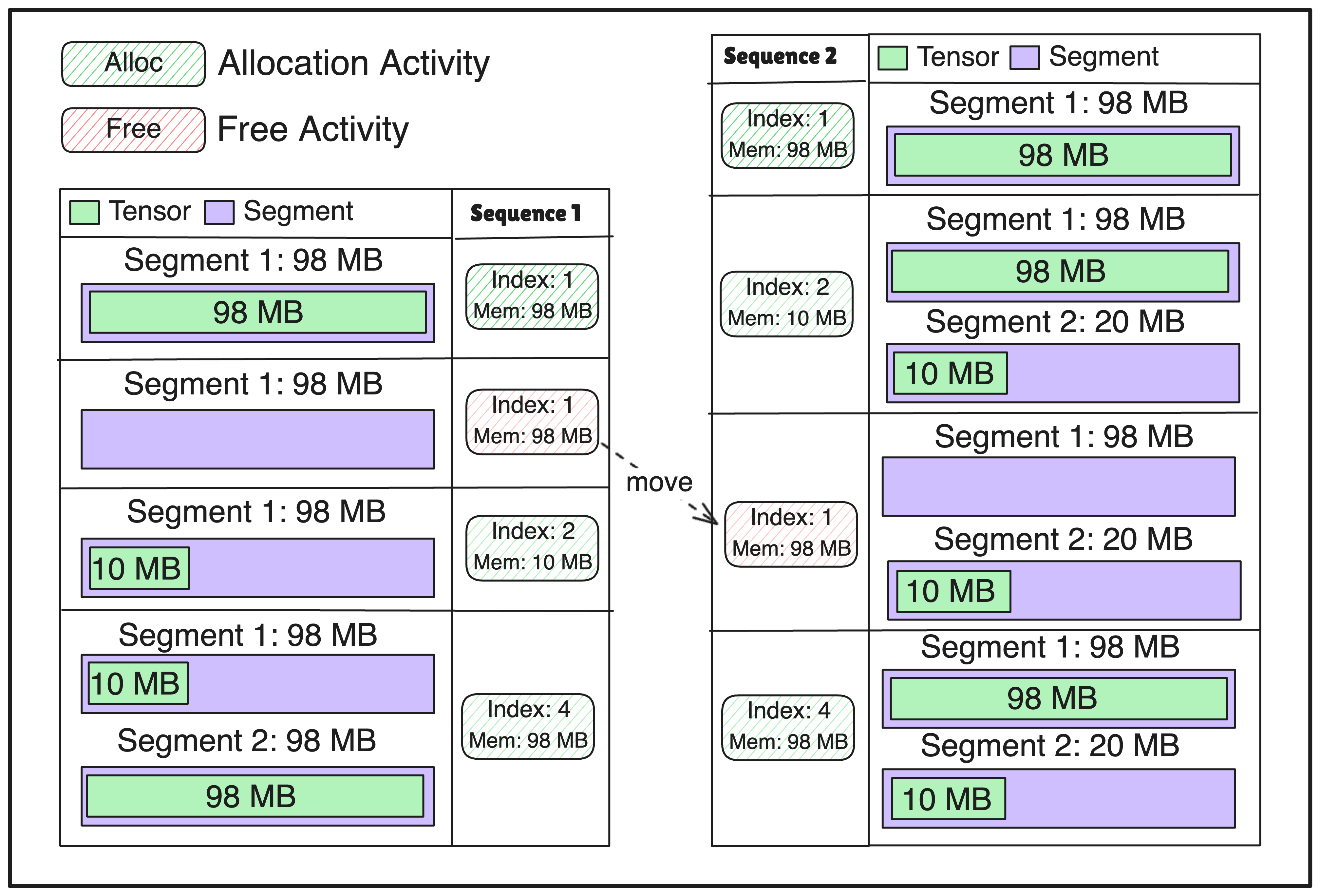}
    \caption{The impact of the sequence of memory activities on GPU memory consumption.}
    \Description{The diagram illustrates two sequences demonstrating how memory operations such as allocation, deallocation, and slight changes in memory operation positions influence memory segments over time.}
    \label{background/fig/memorg-impact-by-sequence}
\end{figure}

Beyond understanding the aforementioned types of memory usage, the temporal sequence of memory allocation and deallocation activities is also critically important in determining peak memory consumption. As shown in Figure~\ref{background/fig/memorg-impact-by-sequence}, merely altering the deallocation timing of one memory block relative to subsequent allocations (Sequence 1 vs. Sequence 2) can dramatically change the peak required GPU segment memory, from 196MB to 118MB, respectively, even when processing identical tensors. This sensitivity highlights the necessity of accurately modeling the dynamic sequence of memory events for reliable peak memory estimation.

\section{Design and Implementation}
\label{design/chapter}
\begin{figure*}[t]
\centering
\includegraphics[width=\linewidth]{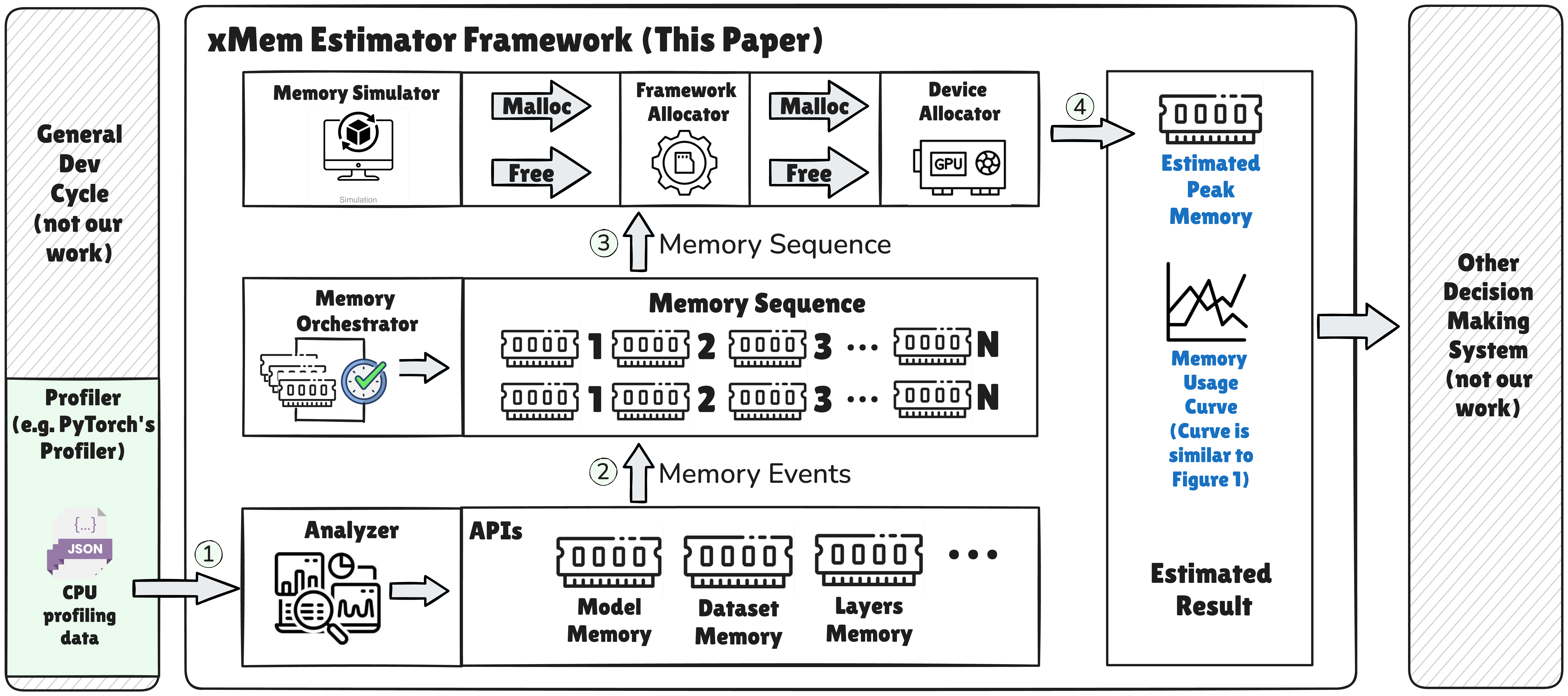}
\caption{Architecture of the \projectname CPU-based memory estimation framework. The pipeline (center) uses CPU profiling data to estimate peak GPU memory via analysis, orchestrator, and simulation.}
\Description{The overall architecture of the \projectname Estimator Framework is illustrated from left to right: general model development cycle -> \projectname Estimator Framework -> results. The framework comprises a data analyzer, memory orchestrator, and simulator. }
\label{design/fig/architecture}
\end{figure*}

\subsection{Architecture} \label{design/chapter/architecture}
The overall architecture of the \projectname estimation framework is presented in Figure~\ref{design/fig/architecture}. The process begins by profiling the initial three iterations\footnote{Persistent states are allocated in the first iteration, with memory stabilizing by iterations 2-3, so three iterations are designed to capture stateful optimizer behavior and stabilized memory.} by default of the target DL training job on the CPU. Crucially, \projectname only requires profiling data from these initial iterations for its analysis, meaning that the DL training job does not need to proceed further for estimation purposes. This Profiling stage captures detailed execution trace data about operator calls, function calls, and memory allocation/deallocation events, producing data in JSON format.

The collected profiling data serve as input to the \projectname estimation framework, which operates as a sequential pipeline of three main components:
\begin{enumerate}
    \item \textbf{Analyzer}: parses raw profiling data, extracts initial memory blocks with CPU lifecycles, filters CUDA-related memory from CPU activities, and attributes filtered events to operators and components like model layers. 
\item \textbf{Memory Orchestrator}: receives the initial memory sequence with CPU-based lifecycles. It refines these timings to more accurately reflect expected GPU memory lifecycles, based on analysis of execution patterns observed in real-world workloads on target GPUs and framework semantics. 
    \item \textbf{Memory Simulator}: replays the orchestrated sequence using a high-fidelity, two-level allocator simulation (DL framework and device interactions) to accurately estimate memory usage over time and mimic target GPU behavior.
\end{enumerate}

The key result produced by the framework is an estimated peak memory required for the job on a target GPU, along with an optional detailed memory usage curve over time. As depicted in Figure~\ref{design/fig/architecture}, these results can be subsequently provided to other decision-making systems such as cluster schedulers, although such systems are outside the scope of this work. As introduced in Section~\ref{introduction/chapter}, the viability of this CPU-only pipeline is based on key observations of modern DL frameworks. In the following subsections, we introduce each \projectname component, using PyTorch \cite{paszke_pytorch_2019} as our primary platform and GPU as the primary device. 
Circled numbers (\eg \circled{1}) indicate key data transfers between system components, as marked in Figure~\ref{design/fig/architecture}.

\subsection{Analyzer}
As the first component of the \projectname pipeline, the Analyzer's main role is to \circled{1} parse raw CPU profiling data and \circled{2} produce filtered GPU-relevant structured memory events, connecting each allocation/deallocation to its originating operation or component. The Analyzer consumes the detailed data generated by PyTorch Profiler~\cite{pytorch_pytorch_2024} during the initial 3--5 iterations of the target DL training job executed on the CPU. For its analysis, \projectname focuses primarily on four event types.
\begin{enumerate}
    \item \textbf{\texttt{python\_function}}: These trace the execution of Python calls, enabling the identification of high-level operations such as specific PyTorch layer invocations, \eg \texttt{Embedding}, \texttt{ReLU}, and \texttt{Linear}. They provide parent-child relationships, thereby forming a call hierarchy.
    \item \textbf{\texttt{user\_annotation}}: These represent markers that signify specific points or phases during the training trace. \projectname harnesses annotations that mark key training loop events, such as \texttt{profiler.step()} for capturing the timeframe of each iteration and \texttt{optimizer.zero\_grad()} for pinpointing gradient clearing actions. These events provide essential information for the Memory Orchestrator.
    \item \textbf{\texttt{cpu\_op}}: These trace the execution of computational kernels \cite{pytorch_cpp_2024} dispatched to the CPU back-end, \eg \texttt{aten::empty}, \texttt{aten::} \texttt{embedding}. They include precise start and end timestamps and often contain sequence numbers that link forward operators to their corresponding backward gradient computations.
    \item \textbf{\texttt{cpu\_instant\_event}}: These capture memory-related activities, primarily recording allocation and deallocation events with associated memory addresses, sizes, and timestamps. A key challenge addressed by the Analyzer is that raw \texttt{cpu\_instant\_event} traces lack explicit linkage back to the specific \texttt{python\_function} and \texttt{cpu\_op} calls that triggered them.
\end{enumerate}

To connect raw \texttt{cpu\_instant\_event} events to their originating operation or component, the Analyzer first reconstructs memory lifecycles. It processes the event stream sequentially, systematically pairing allocation and deallocation events based on address tracking and timing to determine the size, CPU allocation time, and CPU deallocation time for each distinct memory block while correctly handling address reuse. Blocks lacking a deallocation event are considered persistent for the trace duration. Throughout the rest of the paper, the term "memory block" refers specifically to these reconstructed lifecycle entities, unless otherwise specified.

Subsequently, the Analyzer performs hierarchical, time-based attribution using execution windows derived from \texttt{python\_function} and \texttt{cpu\_op} events. A memory block is attributed to a specific operator context if either of the following conditions is satisfied:
\begin{enumerate*}[label=(\roman*)]
    \item the memory block's entire reconstructed lifespan (from allocation to deallocation) falls strictly within that operator's execution time window; or
    \item the memory block is allocated during the operator's execution window but persists beyond the end of the linked high-level component (\eg model layer).
\end{enumerate*}
This operator-centric attribution strategy aims to filter out temporary memory usage allocated during the higher-level script, but not in the operator. These memory blocks are presumed to be more relevant for estimating the requirements of the target GPU.

\begin{figure}[htbp]
\centering
\includegraphics[width=\linewidth]{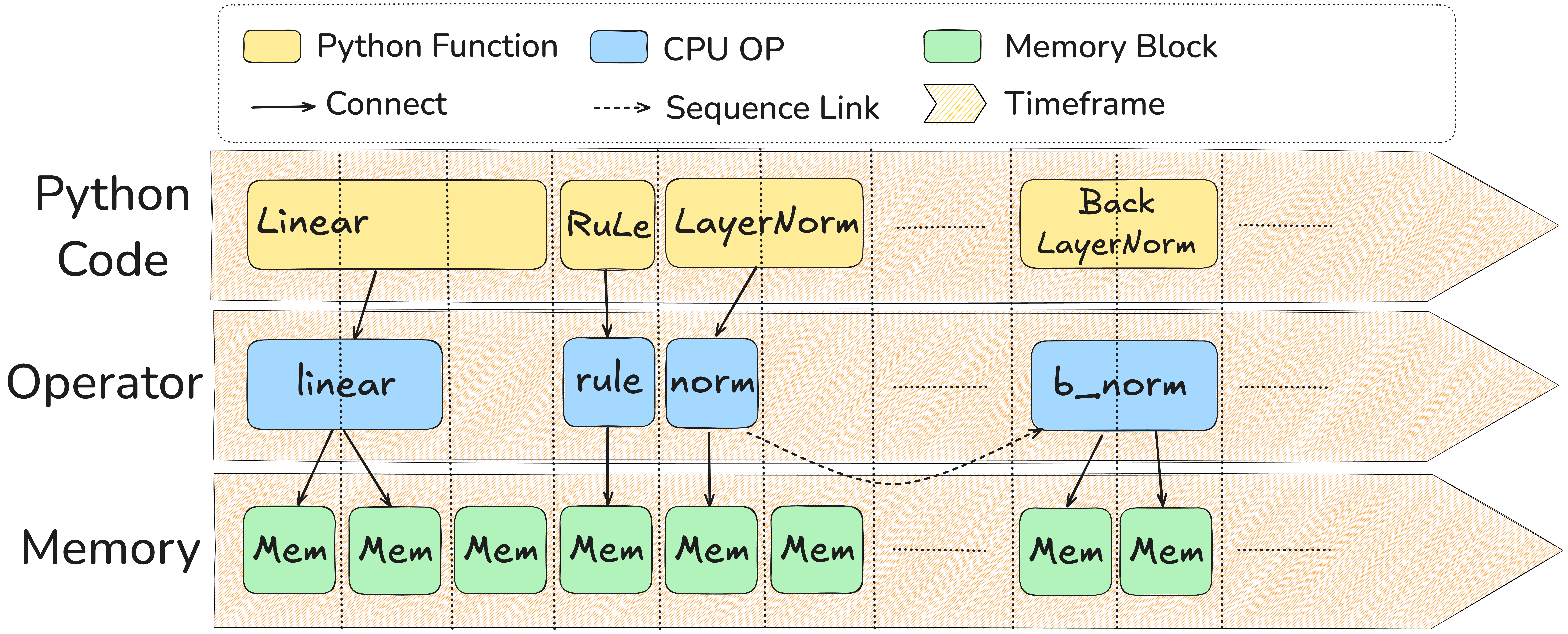}
\caption{A visualization of the time-based hierarchical attribution in \projectname, correlating Python function calls, CPU operator execution time frames, and associated memory blocks.}
\Description{This figure illustrates the hierarchical linking of Python functions, CPU operators, and memory blocks based on execution timing, as used in the \projectname Analyzer.}
\label{design/fig/hierarchical-attribution}
\end{figure}

The Analyzer outputs a structured, temporally ordered sequence of the memory blocks identified through the above process and is verified via the comparison result from Snapshot \cite{pytorch_understanding_2024}. Each memory block record contains its size, initial CPU-based allocation and deallocation timestamps, and association with the originating operator or high-level component, as shown in Figure~\ref{design/fig/hierarchical-attribution}. This temporally ordered sequence of memory blocks provides the foundational input for the next stage, the Memory Orchestrator. 

\subsection{Memory Orchestrator}
The purpose of the Orchestrator is to refine the temporal characteristics of input memory events \circled{2}, tuning the timestamps of key events to more accurately reflect the original memory blocks' lifecycle on the target GPU. This stage is crucial because raw CPU timings may not accurately reflect how long memory blocks are actually required or retained in a GPU. We used a standard training loop code \cite{pytorch_training_2023} to guide the following adjustments.
\begin{enumerate}
    \item \textbf{Model Parameters}: Memory blocks associated with an initial model loading (corresponding to \texttt{model.to(device)}) are marked with a persistent lifecycle, reflecting their presence throughout the analyzed iterations.
    \item \textbf{Batch Data}: Memory blocks linked to loading batch data have lifecycles limited within one training iteration; their deallocation is based on either direct deallocation events (\eg specific \texttt{user\_annotation} event like \texttt{dataloader.\_\_next\_\_}) or iteration boundary markers in the trace.
    \item \textbf{Activations}: For memory blocks corresponding to intermediate activations generated during forward and backward propagation, the Orchestrator largely retains the lifecycle timings determined by the Analyzer. This is because the CPU-derived lifecycles for these intermediate tensors often serve as approximations of their expected lifecycles on GPUs.
    \item \textbf{Gradients}: They are generated during the backward pass, require specific timing adjustments. In a common GPU training loop \cite{pytorch_training_2023}, these gradient tensors persist in memory until the optimizer explicitly clears them via the \texttt{optimizer.zero\_grad()} call. Therefore, the Orchestrator identifies gradient-related memory blocks and adjusts their deallocation timestamps to precisely align with the execution time of the corresponding \texttt{optimizer.zero\_grad()} operation, located using \texttt{user\_annotation} markers in the trace. 
    \item \textbf{Optimizer}: Potential optimizer memory blocks are identified during the \texttt{optimizer.step()} phase by filtering for blocks whose sizes match model parameters. Since stateful optimizers such as Adam \cite{kingma_adam_2014} allocate persistent states mainly in the first iteration (unlike SGD \cite{ruder_overview_2016} with minimal overhead), analyzing at least two iterations is crucial. This is because the peak memory usage in the second iteration is further analyzed based on the persistent memory blocks allocated in the first iteration.
\end{enumerate}
The Memory Orchestrator produces a refined memory sequence\footnote{Although the orchestrated sequence has been refined to match the expected sequence in the target GPU, its effectiveness is limited by substantial underlying behavioral divergences. A key challenge for CPU-based estimation accuracy is that fundamental differences in operator implementation and optimization strategies across backends significantly degrade the prediction accuracy.} \circled{3} to reflect the anticipated sequence in the GPU, then feeds it as input for the Memory Simulator.

\subsection{Simulator}

The final pipeline stage, the Memory Simulator, replays the orchestrated memory sequence \circled{3}. It employs a two-level simulation of both the target GPU's memory allocator and the DL framework's internal allocator. Precisely simulating the allocator's behavior is crucial because simply summing tensor sizes overlooks complex mechanisms like caching and fragmentation (as discussed in Section~\ref{background/chapter/memory}), enabling \projectname to accurately track the dynamic memory footprint and estimate peak GPU memory consumption \circled{4}.

To capture these dynamics, the \projectname Simulator implements key techniques of such a caching allocator:
\begin{enumerate*}[label=(\roman*)]
    \item \textbf{Round up}: All memory request sizes are rounded up to the nearest hardware-required multiple (\eg 512 bytes for CUDA operations \cite{pytorch_cuda_2024}), reflecting how allocators often handle physical memory constraints.
    \item \textbf{Segment}: The Simulator involves the concept of Segments, larger memory blocks that the framework's allocator requests from the GPU. The strategy\footnote{The segment's allocation strategy follows the PyTorch Official implementation \cite{pytorch_github-pytorch_2024}} for determining segment sizes is incorporated to mimic how allocators balance allocation speed with potential fragmentation.
    \item \textbf{Algorithm}: To accurately reproduce the behavior of both the DL framework allocator (\eg PyTorch's CUDACachingAllocator \cite{pytorch_github-pytorch_2024, pytorch_cuda_2024}) and the GPU allocator \cite{calderon_gmai_2020}, the Simulator primarily employs the Best Fit with Coalescing (BFC) algorithm \cite{hasan_study_2005}. This BFC implementation involves searching for available free blocks within segments, splitting blocks when an exact match is unavailable, and merging adjacent free blocks upon deallocation to closely mimic real-world fragmentation and coalescence patterns.
    \item \textbf{Caching Behavior}: Upon a deallocation in the event sequence, the Simulator caches the block within its segment for reuse. New segments are requested from the GPU only if this cache is insufficient for an allocation. Cached blocks persist until the framework allocator needs more memory, but the device indicates an OOM error.
    \item \textbf{OOM}: This two-level simulation, modeling both the framework and device-level allocator, captures the complete memory management chain. An OOM condition is signaled only when memory requests fail at both simulated allocator levels, even after attempting the reclamation of cached segments.
\end{enumerate*}

\begin{figure}[htbp]
    \centering
    \includegraphics[width=\columnwidth]{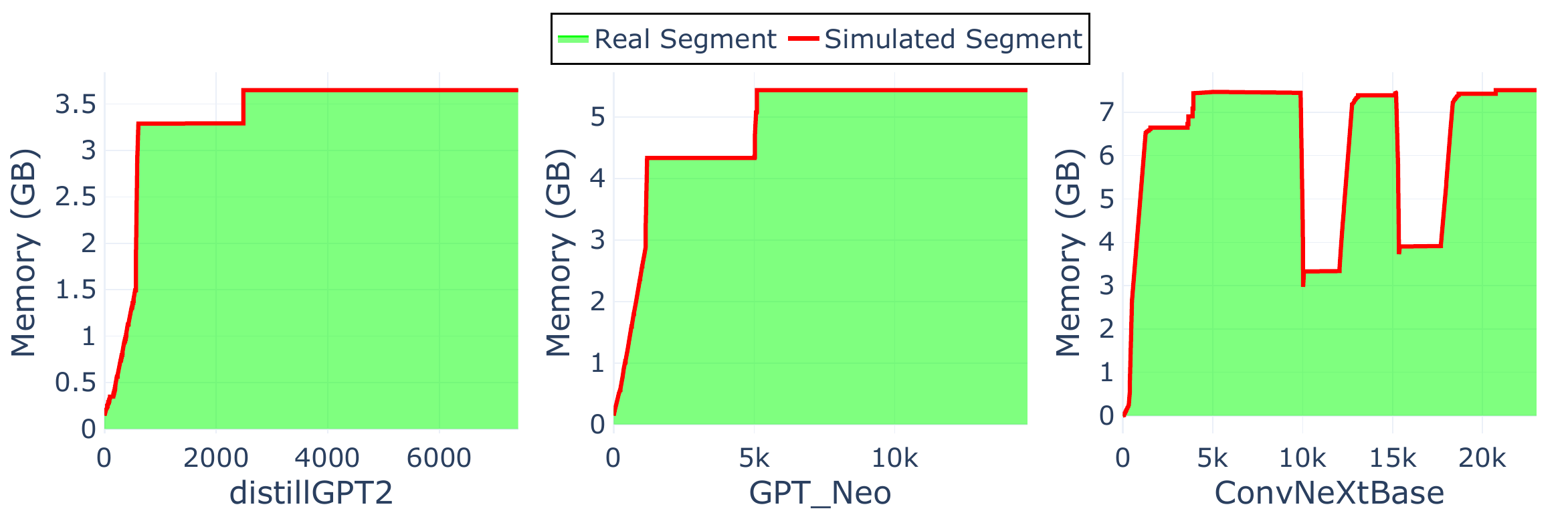}
    \caption{Comparison of actual GPU memory segment usage (from PyTorch Snapshot Profiler \cite{pytorch_understanding_2024}) with xMem's simulated segment usage for three DL models, validating the Simulator's accuracy.}
    \Description{This figure displays three plots comparing actual GPU memory segment usage (green area) with \projectname simulated segment usage (red line) over time for the ConvNeXtBase, distillgpt2, and Cerebras-GPT-111M models, showing a close match between the two.}
    \label{design/fig/allocator-assessment}
\end{figure}

Regarding the above techniques, the Simulator processes the orchestrated memory event sequence chronologically. For each allocation event, it attempts to secure memory via the simulated two-level allocator. Correspondingly, for each deallocation event, it marks the block as free within the simulated allocator's pools, which may trigger coalescing. By tracking the total size of the segments allocated from the GPU at each point in the event sequence, \projectname determines the time series of memory usage for Tensors and Segments. The Estimated Peak Memory is then identified as the maximum value in this time series, and the full series can optionally be output as a memory usage curve for detailed analysis. To verify the accuracy of this simulation approach, we used actual PyTorch allocator behavior obtained using PyTorch's snapshot profiler \cite{pytorch_understanding_2024} compared to the Simulator result. These comparative results are shown in Figure~\ref{design/fig/allocator-assessment}.

\section{Evaluation}
\label{evaluation/chapter}
\begin{table}[ptbp]
    \caption{Notations used in the evaluation experiments.}
    \begin{center}
    \rowcolors{2}{}{stripe}
    \begin{tabular}{>{\centering\arraybackslash}m{1.1cm} m{6.6cm}}
    \toprule
    \textbf{Notation} & \multicolumn{1}{c}{\textbf{Definition}} \\
    \midrule
    $j$ & One test configuration, including model, optimizer, \etc \\
    $i$ & Verification Round: 1\textsuperscript{st} or 2\textsuperscript{nd}\\
    $d$ & The index of the target GPU device, $d \in \{0,1\}$ \\
    $e$ & The estimator\\
    $N$ & The number of performed runs\\
    $M^{\text{init}}_{d}$ & The amount of memory used on device $ d$ for the duration of the experiment \\
    $M^{\text{fm}}$ & The amount of memory used by a DL Framework. It is generally a constant amount of memory during all experiments \\
    $M_{d}^{\text{max}}$ & The memory capacity of device $d$ \\
    $M^{\text{peak}}_{jid}$ & The peak memory usage as recorded by NVML during training with configuration $j$ on device $d$ at the $i$\textsuperscript{th} validation \\
    $\hat{M}^{\text{peak}}_{jde}$ & The peak memory usage as predicted by estimator $e$ using configuration $j$ and memory capacity of device $d$ \\
    $M^{\text{save}}_{jde}$ & The memory conserved by estimator $e$ with configuration $j$ on device $d$ \\
    $\text{avgM}^{\text{save}}_{e}$ & The average memory conserved by running the estimator $e$\\
    $\hat{\text{OOM}}_{jde}$ & Boolean prediction of OOM occurrence by estimator $e$ using configuration $j$ on device $d$ \\
    $\text{OOM}_{jd1}$ & Boolean indicating actual OOM occurrence during training with configuration $j$ on device $d$ at 1\textsuperscript{st} validation \\
    $\text{OOM}_{jde2}$ & Boolean indicating actual OOM occurrence during training with configuration $j$ under maximum runnable GPU memory estimated by estimator $e$ on device $d$ at 2\textsuperscript{nd} validation \\
    $C_{jide}$ & Boolean indicating whether the prediction $\hat{\text{OOM}}_{jde}$ matches the actual $\text{OOM}_{jdi}$ at the $1$\textsuperscript{st} validation. In the $2$\textsuperscript{nd} validation, $C_{jde1}$ is utilized to further assess its estimation status \\
    $\text{error}_{jide}$ & The error of $M^{\text{peak}}_{jid}$ relative to $\hat{M}^{\text{peak}}_{jde}$ \\
    $\tilde{\text{error}}_{jide}$ & The median of a set of $\text{error}_{jide}$  \\
    $P_{jie}$ & Probability of estimation failure by estimator $e$ with configuration $j$ at the $i$\textsuperscript{th} validation \\
\bottomrule
    \end{tabular}
    \label{evaluation/table/notations}
    \end{center}
\end{table}

This section presents a comprehensive empirical evaluation of \projectname to validate its effectiveness in accurately estimating peak GPU memory consumption for DL training jobs. We conducted experiments across various GPU types, including A100, GeForce RTX 4060, and GeForce RTX 3060, and compared \projectname with SOTA solutions representing diverse approaches to memory estimation. 
The evaluation aims to answer the following key research questions: \begin{enumerate}
    \item \textbf{RQ1 (Accuracy)}: How accurate are the peak memory estimations provided by \projectname? (Section~\ref{evaluation/chapter/rq/accurate})
    \item \textbf{RQ2 (Reliability)}: How reliably does \projectname prevent OOM errors? (Section~\ref{evaluation/chapter/rq/reliably})
    \item \textbf{RQ3 (Memory Conservation)}: How much GPU memory can potentially be conserved by using \projectname's estimates? (Section~\ref{evaluation/chapter/rq/memorysaving})
    \item \textbf{RQ4 (Overhead)}: What is the run-time overhead of \projectname? (Section~\ref{evaluation/chapter/rq/runtime})
\item \textbf{RQ5 (Scalability)}: How accurately does \projectname estimate peak memory for larger Transformer models in high-end GPUs, such as the A100? (Section~\ref{evaluation/chapter/rq/largermodel})
\end{enumerate}

\subsection{Evaluation Methodology}
The notations used in this study are summarized in Table~\ref{evaluation/table/notations}.

\subsubsection{Baselines}
We compare \projectname against three state-of-the-art GPU memory estimators that employ distinct methodologies:
\begin{enumerate}
    \item \textbf{DNNMem \cite{gao_estimating_2020}}: Represents static analysis approaches, which combine computational graph analysis with the simulation of a basic BFC allocator. As the source code is not publicly available, we reproduced DNNMem based on the descriptions and algorithms presented in the original paper.
    \item \textbf{SchedTune \cite{albahar_schedtune_2022}}: Represents machine learning-based approaches, utilizing pre-trained models based on model features and hardware characteristics to estimate memory needs.
    \item \textbf{LLMem \cite{kim_llmem_2024}}: Represents approaches relying on direct GPU interaction, specifically designed for estimating memory for Large Language Model fine-tuning through targeted GPU execution measurements.
\end{enumerate}
For consistency, all evaluation plots use the following color scheme: {\color{exp-xMem}{blue}} for \projectname, {\color{exp-DNNMem}red} for DNNMem, {\color{exp-SchedTune}green} for SchedTune, and {\color{exp-LLMem}purple} for LLMem. To establish the ground truth peak memory values used in calculating MRE, we sampled the total allocated GPU memory at 1 ms intervals during actual training runs using the NVIDIA Management Library (NVML) \cite{nvidia_corporation_nvml-nvidia_2024}. We designated the maximum value recorded across all samples during each run as the ground truth peak.

\subsubsection{Workload} \label{eva/chapter/workload}

\begin{table}[btp]
\caption{DL models (CNN and Transformer) and optimizers used in the evaluation. `$*$' denotes models used for RQ5 only.}
\Description{Table listing the Convolutional models (VGG, ResNet, MobileNetV2/V3, MnasNet, RegNetX/Y, ConvNeXt) and Transformer models (DistilGPT2, GPT2, T5, GPT-Neo, OPT, Cerebras-GPT) with release years, alongside the five optimizers (AdamW, Adam, SGD, RMSprop, Adagrad) used in the evaluation experiments.}
\begin{center}
\begin{tabular}{cc|c}
\toprule
\textbf{Convolutional Model} & \textbf{Year} & \textbf{Optimizers} \\
\midrule
VGG (16 \& 19) \cite{simonyan_very_2014} & 2014 &   \\
ResNet (101 \& 152) \cite{he_deep_2016} & 2016 &   AdamW \\
MobileNetV2 \cite{sandler_mobilenetv2_2018} & 2018 & Adam \\
MobileNetV3 (Small \& Large) \cite{howard_searching_2019} & 2019 & SGD \\
MnasNet \cite{tan_mnasnet_2019} & " & RMSprop  \\
RegNetX (400MF) \cite{radosavovic_designing_2020} & 2020 & Adagrad  \\
RegNetY (400MF) \cite{radosavovic_designing_2020} & " &  \\
ConvNeXt (Tiny \& Base) \cite{liu_convnet_2022} & 2022 &  \\
\midrule
\textbf{Transformer Model} & & \\
\midrule
DistilGPT2 \cite{sanh_distilbert_2019} & 2019 & \\
GPT2 \cite{radford_language_2019} & " & \\
T5 (small \& base) \cite{2020t5} & 2020 & \\
GPT-Neo (125M) \cite{liu_convnet_2022} & 2022 & AdamW \\
OPT (125m \& 350m) \cite{zhang_opt_2022} & " & Adam \\
Cerebras-GPT (111M) \cite{dey_cerebras-gpt_2023} & 2023 &  SGD\\
Pythia (1b) \cite{biderman_pythia_2023} & " & Adafactor \\
$*$ Llama-3.2 (3B-Instruct) \cite{grattafiori_llama_2024} & 2024 &\\
Qwen3 (0.6b) \cite{yang_qwen3_2025} & 2025 & \\
$*$ DeepSeek-R1 (Distill-Qwen-1.5B) \cite{deepseek-ai_deepseek-r1_2025} & " & \\
$*$ Qwen3 (4B) \cite{yang_qwen3_2025} & " & \\
\bottomrule
\end{tabular}
\label{evaluation/table/model}
\end{center}
\end{table}

To ensure a comprehensive assessment, our evaluation employs a diverse set of workloads. For RQ1-RQ4, we utilize 22 different DL models (without $*$ symbol), detailed in Table~\ref{evaluation/table/model}, encompassing widely-used architectures such as CNNs \cite{krizhevsky_imagenet_2017} and Transformers \cite{vaswani_attention_2017}. Optimizer selection and batch size ranges are tailored to each model architecture: for CNN models, we employ SGD, Adam, AdamW, RMSprop, and Adagrad, with the batch sizes ranging from 200 to 700 with a step of 100; for Transformer models, we use SGD, Adafactor, Adam, and AdamW (excluding RMSprop and Adagrad due to their less common usage in this context), varying the batch size from 5 to 55 with a step of 5. In addition, we used a smaller batch size, \eg 1 to 8 with a step of 1, for Qwen3 and Pythia models due to their higher parameter counts. Applying the same range of batch sizes across all models would be self-defeating, as some of the models only experience OOM errors during all experiments. Furthermore, to assess the impact of code variations on memory usage, we additionally involved different placements of the \texttt{optimizer.zero\_grad()} call in Monte Carlo simulations, reflecting real coding patterns that influence memory behavior (see Figure~\ref{motivation/fig/zero-out-memory-change}).

For RQ5, a distinct set of 3 Transformer models was used, as highlighted in Table~\ref{evaluation/table/model} with a $*$ symbol. These models were chosen to test the scalability of \projectname on larger Transformer models. Additionally, given that this experiment focuses on accuracy, we ensured all tests were conducted within GPU memory limits to guarantee that each run could provide a valid MRE. To achieve this, we solely employed the SGD and Adafactor optimizers because all models could be trained without encountering an OOM error. Each trial was repeated five times, and the batch size was set to 1.

\subsubsection{Environment Setup}
Experiments for RQ1-RQ4 were conducted on a server equipped with a 24-core Intel i9 CPU and 128GB of RAM. For GPU-related ground truth measurements and baseline executions requiring a GPU, we utilized two NVIDIA GPUs: a GeForce RTX 3060 with 12GB of memory and a GeForce RTX 4060 with 8GB of memory. To ensure consistency and reproducibility, the software environment was managed using Docker containers built on official PyTorch images \cite{pytorch_docker_2025}, including the necessary Python dependencies. As many conflict issues exist among dependencies required by baselines and xMem, \projectname and DNNMem were evaluated using \texttt{2.6.0-cuda12.4-cudnn9-devel} tag; SchedTune utilized \texttt{2.3.1-cuda12.1-cudnn8-devel} tag, whereas LLMem required \texttt{2.0.1-cuda11.7-cudnn8-devel} tag. During experiments, each GPU was dedicated exclusively to a single run to prevent interference.

For RQ5, experiments were conducted using Google CoLab Pro instances, providing access to NVIDIA A100 GPUs with 40GB of GPU memory. This distinct environment was chosen to assess the scalability and performance of \projectname on high-end accelerator hardware with larger model sizes. Since package conflict issues were encountered with SchedTune and LLMem in this CoLab environment, \projectname only compares with DNNmem. The ground truth peak memory on the A100 was also obtained using the same NVML-based sampling methodology.

\subsubsection{Experimental Design} \label{evaluation/chapter/experimental-design}
To strictly assess each estimator $e$ for a given test configuration j (model, optimizer, batch size, \texttt{zero\_grad} placement) on a device $d$, we employ a two-round validation process. \textbf{First (Initial Validation, $i=1$)}, we compare an estimated $\hat{OOM}_{jde}$, expressed as Eq.~\ref{evaluation/equation/estimation-oom}, against the actual $OOM_{jd1}$ of running the job with full device memory, while recording $M^\text{peak}_{jd1}$ and $M^{fm}$. \textbf{Second (Subsequent Validation, $i=2$)}, we execute the job with the same configuration again when only $C_{jde1}=1 \land OOM_{jd1}=0$, this time using $M^\text{init}_d + M^{fm} + \hat{M}^{\text{peak}}_{jde}$ as the maximum runable memory, while recording $OOM_{jde2}$ and $M^\text{peak}_{jd2}$. 

\begin{equation}
    \label{evaluation/equation/estimation-oom}
    \hat{\text{OOM}}_{jde} = [\hat{M}^{\text{peak}}_{jde} > M_{d}^{\text{max}}]
\end{equation}

The success or failure in this second round is crucial in determining the MRE, PEF, and MCP metrics, reflecting the practical reliability and the implications of resource savings of the estimate. This validation is applied in two main experimental settings:
\begin{enumerate}
    \item \textbf{ANOVA}: For comprehensive runs, we systematically selected specific configurations by pairing all models with their applicable optimizers and batch sizes as defined in Section~\ref{eva/chapter/workload}. We repeated each configuration run five times on the GeForce RTX 3060 GPU, using results from a total of 3903 runs to analyze error distributions and reliability under controlled conditions, mainly to address RQ1, RQ2, and RQ5.
    \item \textbf{Monte Carlo Simulations \cite{rubinstein_simulation_2017}}: We conducted 1306 runs where each training configuration (all pairs are mentioned in Section~\ref{eva/chapter/workload} and two target GPUs) was randomly selected. We utilize the nature of the Monte Carlo, simulating the randomness and uncertainty of reality, to assess the overall performance (RQ1-RQ4) of the estimators, especially the memory conservation potential (RQ3).

\end{enumerate}

\begin{figure*}[t]
\centering
\begin{subfigure}{\columnwidth}
    \includegraphics[width=\textwidth]{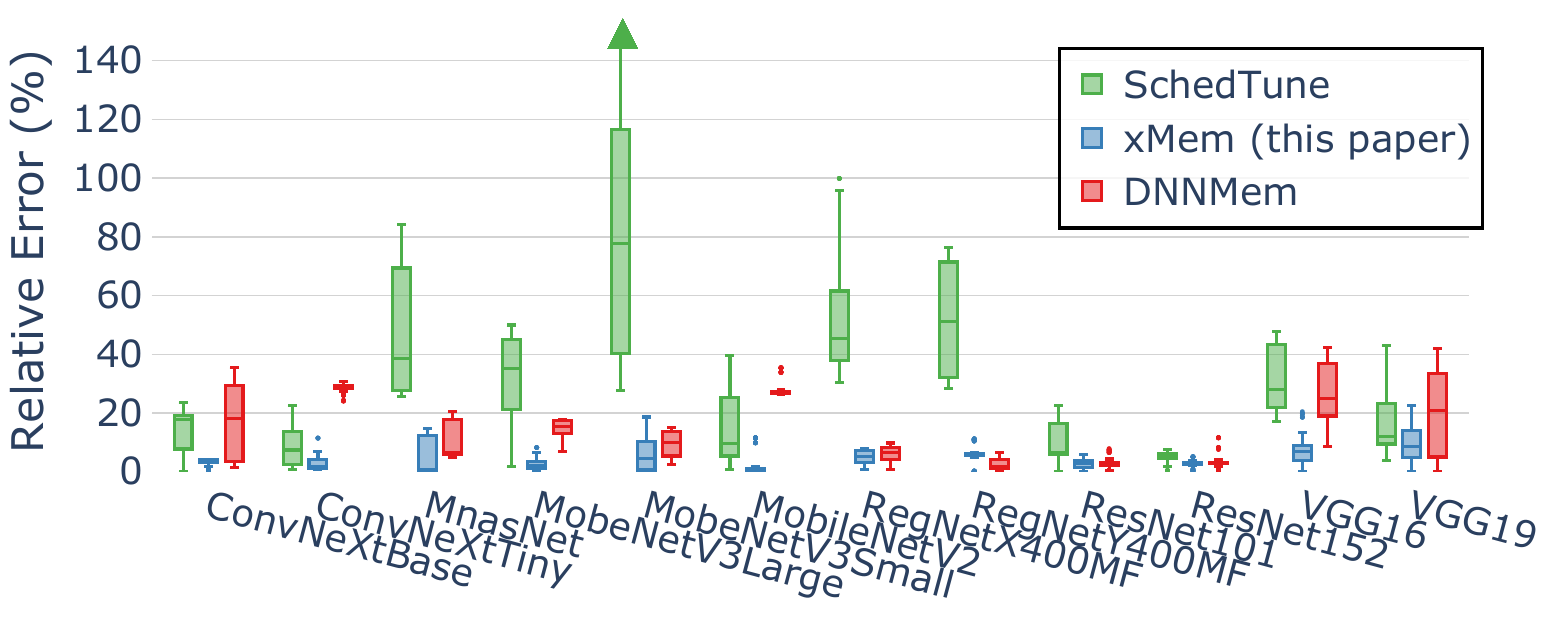}
    \caption{CNN Models (ANOVA)}
    \Description{A box plot chart displaying the Median Relative Error (MRE) percentages for peak memory estimations across multiple Convolutional Neural Network (CNN) models from ANOVA experiments. The chart compares the performance of three estimators: \projectname (typically shown in blue), DNNMem (typically red), and SchedTune (typically green). Each model on the x-axis has a set of box plots representing the MRE distribution for these estimators.}
    \label{evaluation/fig/MRE/RQ1/CNN}
\end{subfigure}
\hfill \begin{subfigure}{\columnwidth}
    \includegraphics[width=\linewidth]{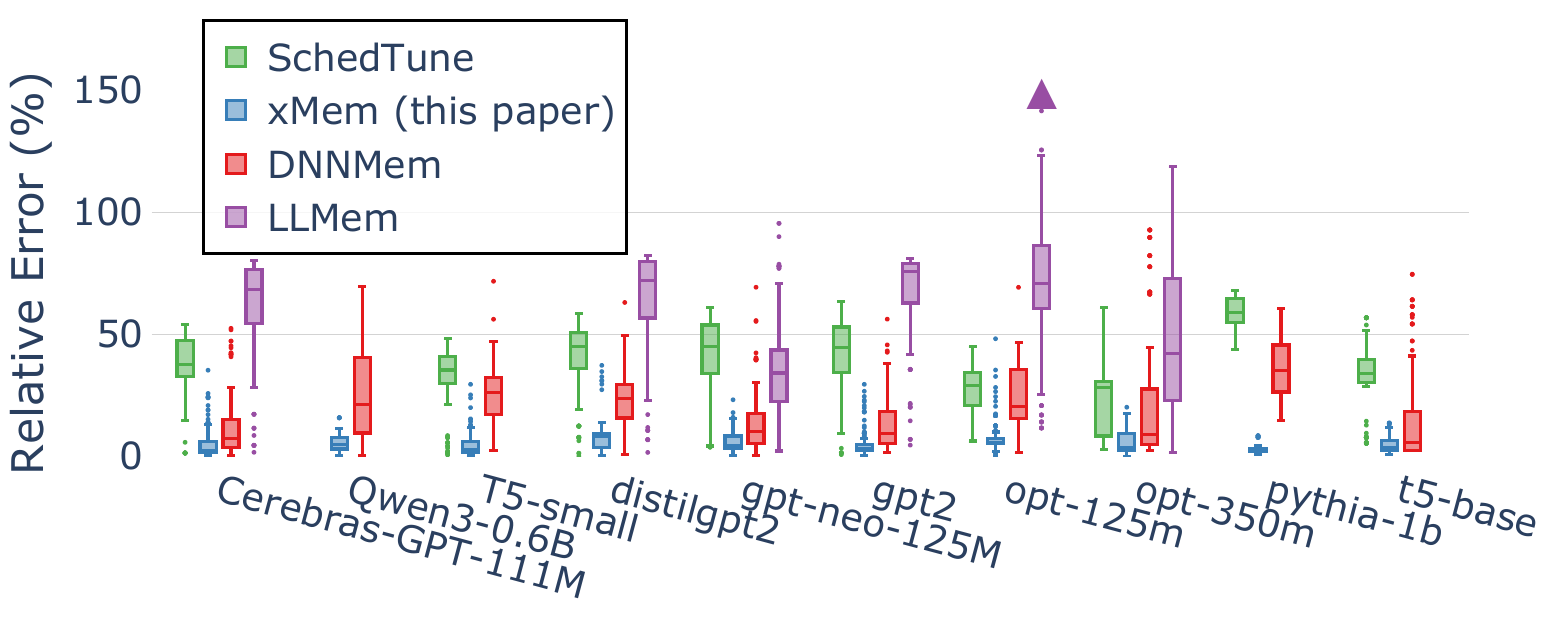}
      \caption{Transformer Models (ANOVA)}
      \Description{A box plot chart displaying the Median Relative Error (MRE) percentages for peak memory estimations across multiple Transformer models from ANOVA experiments. The chart compares the performance of four estimators: \projectname (typically shown in blue), DNNMem (typically red), SchedTune (typically green), and LLMem (typically purple). Each model on the x-axis has a set of box plots representing the MRE distribution for these estimators.}
      \label{evaluation/fig/MRE/RQ1/Transformer}
\end{subfigure}

\begin{subfigure}{\columnwidth}
    \includegraphics[width=\textwidth]{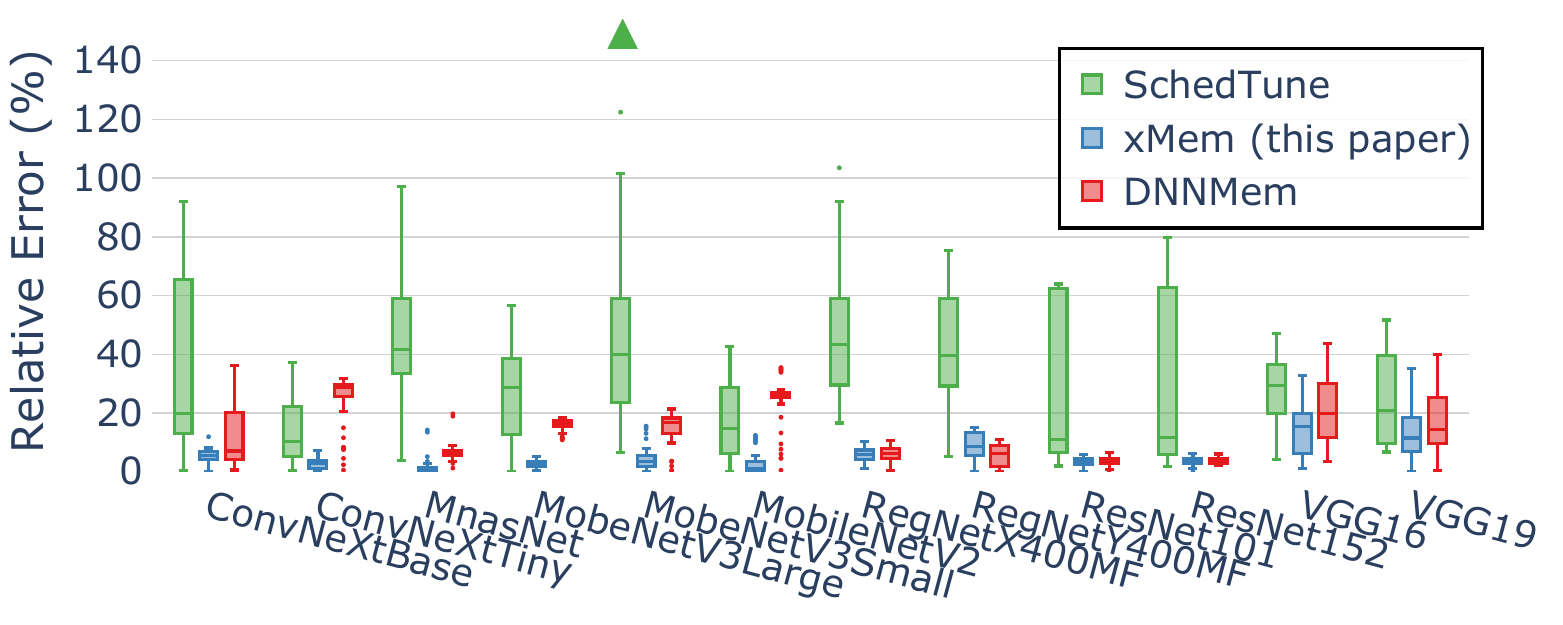}
    \caption{CNN Models (Monte Carlo)}
    \Description{A box plot chart displaying the Median Relative Error (MRE) percentages for peak memory estimations across multiple Convolutional Neural Network (CNN) models from Monte Carlo experiments. The chart compares the performance of three estimators: \projectname (typically shown in blue), DNNMem (typically red), and SchedTune (typically green). Each model on the x-axis has a set of box plots representing the MRE distribution for these estimators.}
    \label{evaluation/fig/MRE/RQ1/CNN/MC}
\end{subfigure}\hfill \begin{subfigure}{\columnwidth}
    \includegraphics[width=\linewidth]{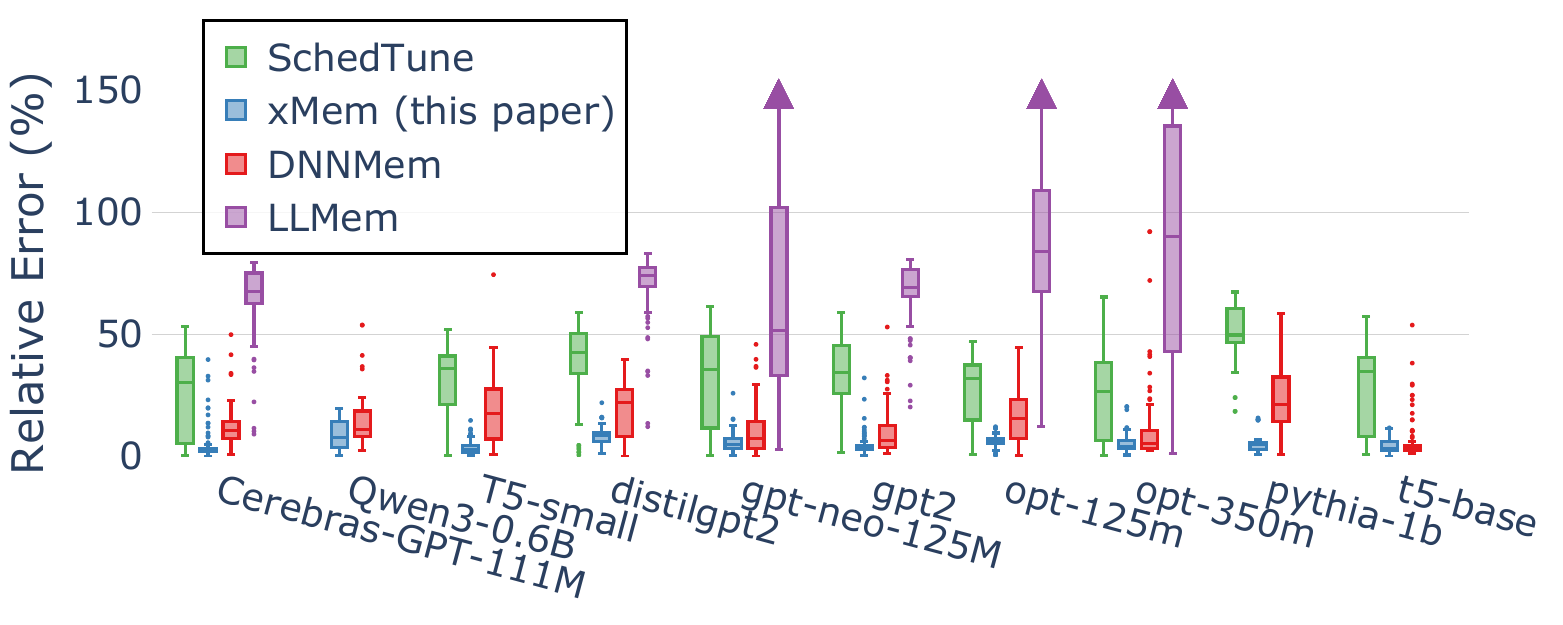}
      \caption{Transformer Models (Monte Carlo)}
      \Description{A box plot chart displaying the Median Relative Error (MRE) percentages for peak memory estimations across multiple Transformer models from Monte Carlo experiments. The chart compares the performance of four estimators: \projectname (typically shown in blue), DNNMem (typically red), SchedTune (typically green), and LLMem (typically purple). Each model on the x-axis has a set of box plots representing the MRE distribution for these estimators.}
      \label{evaluation/fig/MRE/RQ1/Transformer/MC}
\end{subfigure}
\caption{Comparison of the MRE in peak memory estimation across different model architectures. Lower MRE indicates higher accuracy (An absent box plot in a specific model means that the estimator does not support that model.). The symbol $\blacktriangle$ denotes outliers with an MRE exceeding 150\%.}
\Description{A box plot chart displaying the Median Relative Error (MRE) percentages for peak memory estimation across architectures.}
\label{evaluation/fig/MRE/RQ1}
\end{figure*}

\subsubsection{Metrics} \label{evaluation/chapter/metrics}
We evaluate the estimators using the following:
\begin{enumerate}
\item  \textbf{MRE}: This represents a typical (median) percentage by which an estimated peak memory differs from the ground truth value, as expressed in Eq.~\ref{evaluation/equation/median-error}. Since ground truth cannot be measured when a real OOM ($OOM_{jd1}=1$), happens,  only jobs matched a condition, $OOM_{jd1}=0$, are selected for this statistic. Additionally, $\text{error}_{jde2}$ is used when $OOM_{jde2}=0$; otherwise, $\text{error}_{jde1}$ is utilized.
    \item \textbf{PEF}: This represents a probability, across multiple runs, that an estimated memory did not pass our two-round validation check (Section~\ref{evaluation/chapter/experimental-design}), expressed as Eq.~\ref{evaluation/equation/failed-estimation-probability}. To rigorously assess performance in our evaluations, we utilize $P_{je2}$ as a key metric. A lower probability indicates greater reliability.
    \item \textbf{MCP}: The average GPU memory is saved per time by using the estimator. It is expressed as Eq.~\ref{evaluation/equation/estimator-memory-average-saving}, which averages the per-run savings determined by Eq.~\ref{evaluation/equation/memory-savings}. This metric penalizes unreliable estimations to reflect the practical cost of estimation failure: if an $\hat{M}^{\text{peak}}_{jde}$ leads to an OOM failure (determined by $C_{jde2}$), the $\hat{M}^{\text{peak}}_{jde}$ (which would have been allocated and thus wasted) is deducted from the overall memory saved.
    \item \textbf{Runtime}: This measures the time required for each method to estimate a single peak memory for a given configuration.
\end{enumerate}

\begin{equation}
    \text{error}_{jide} = 
    \frac{\| \hat{M}^{\text{peak}}_{jde} - M^{\text{peak}}_{jid} \|}{M^{\text{peak}}_{jid}}, OOM_{jd1} = 0
    \label{evaluation/equation/relative-error}
\end{equation}

\begin{equation}
    \label{evaluation/equation/median-error}
\tilde{\text{error}}_{jide} (MRE) = \text{Median} \left( \begin{cases} error_{jde2} & \text{if } OOM_{jde2} = 0 \\ error_{jde1} & \text{if } OOM_{jde2} \neq 0 \end{cases} \right)
\end{equation}

\begin{equation}
    \label{evaluation/equation/1st-correctness-estimation}
    C_{jde1} = [\hat{\text{OOM}}_{jde} = \text{OOM}_{jd1}]
\end{equation}
\begin{equation}
    \label{evaluation/equation/correctness-estimation/2nd}
    C_{jde2} = [C_{jde1}=1 \land (\text{OOM}_{jde2} = 0 \lor \text{OOM}_{jd1}=1)]
\end{equation}
\begin{equation}
    \label{evaluation/equation/failed-estimation-probability}
     P_{jie} (PEF)= \frac{N-\sum_{n=1}^{N} C_{jide}}{N}
\end{equation}

\begin{equation}
    \label{evaluation/equation/memory-savings}
    M^{\text{save}}_{jde} = 
    \begin{cases} 
    M^{\text{max}}_d - \hat{M}^{\text{peak}}_{jde}, &   C_{jde1}=1 \land \text{OOM}_{jde2} = 0 \\
    M^{\text{max}}_d, &   C_{jde1}=1 \land \text{OOM}_{jd1}=1 \\
    -M^{\text{max}}_d, &  otherwise \\
    \end{cases}
\end{equation}
\begin{equation}
    \label{evaluation/equation/estimator-memory-average-saving}
    \text{avgM}^{\text{save}}_{e} (MCP) = \frac{\sum_{n=1}^{N} M^{\text{save}}_{jde}}{N}
\end{equation}

\subsection{RQ1: How accurate are the peak memory estimations?} \label{evaluation/chapter/rq/accurate}
To answer RQ1, we primarily analyze the MRE, namely $\tilde{\text{error}}_{jide}$, of peak memory estimations. A lower MRE indicates higher accuracy. We present results from the comprehensive runs, which were conducted on the GeForce RTX 3060 GPU.

Figure~\ref{evaluation/fig/MRE/RQ1/CNN} illustrates the MRE distributionsin ANOVA for CNN models, each box representing around 150 samples. Across the 12 CNN models, \projectname consistently demonstrates superior accuracy, with MRE approximately 3\%. For instance, \projectname achieves a MRE of approximately 1.8\% for ConvNeXtTiny and around 0.2\% for MobileNetV2. In most cases, \projectname's MRE is below 5\%, with only four models (VGG16, VGG19, RegNetX400MF, and RegNetY400MF) showing a MRE slightly above this, but still generally below 10\%. In contrast, DNNMem exhibits higher MREs, generally ranging from 10\% to 30\% for 7 of 12 models, with wider interquartile ranges (IQRs) indicating greater variability. SchedTune consistently shows the highest MREs and the largest variance, with MRE often exceeding 20\% for over half of CNN models. LLMem was not evaluated for CNN models as it is specifically designed for Transformer architectures. The consistently tight error distribution of \projectname around 5\% highlights its robustness in accuracy for CNN workloads.

For Transformer models in ANOVA, the MRE distributions are presented in Figure~\ref{evaluation/fig/MRE/RQ1/Transformer}, each box representing around 160 samples for Pythia and Qwen, and 220 samples for the rest of the models. \projectname again shows outstanding performance across the 10 models, with a MRE approximately 4\%. Its MREs are generally low, for example, approximately 2.3\% for Cerebras-GPT-111M, 7.8\% for Qwen3, and 4.8\% for Pythia-1b. DNNMem struggles with Transformer models, displaying MREs often between 10\% and 30\% (\eg for t5-small and distilgpt2), indicating its static analysis is less effective. SchedTune continues to exhibit high MREs, but less variance compared to the ANOVA experiment with CNN models. LLMem, specifically designed for CausalLM, exhibits the highest error, with an outlier exceeding 150\%.

The results from the Monte Carlo simulations are shown in Figures~\ref{evaluation/fig/MRE/RQ1/CNN/MC} and \ref{evaluation/fig/MRE/RQ1/Transformer/MC} (no guaranteed samples per box due to the nature of the Monte Carlo simulation). The simulation covers random configurations across two GPU types, further corroborating these findings. When aggregating MREs from these simulations, \projectname consistently maintains the lowest overall MRE (around 4\% for both CNN and Transformer) and the most concentrated error distribution, demonstrating its robustness across a wide variety of models, optimizers, batch sizes, and code structure variations.

\begin{figure}[htb]
\centering
\begin{subfigure}{0.495\columnwidth}
    \includegraphics[width=\textwidth]{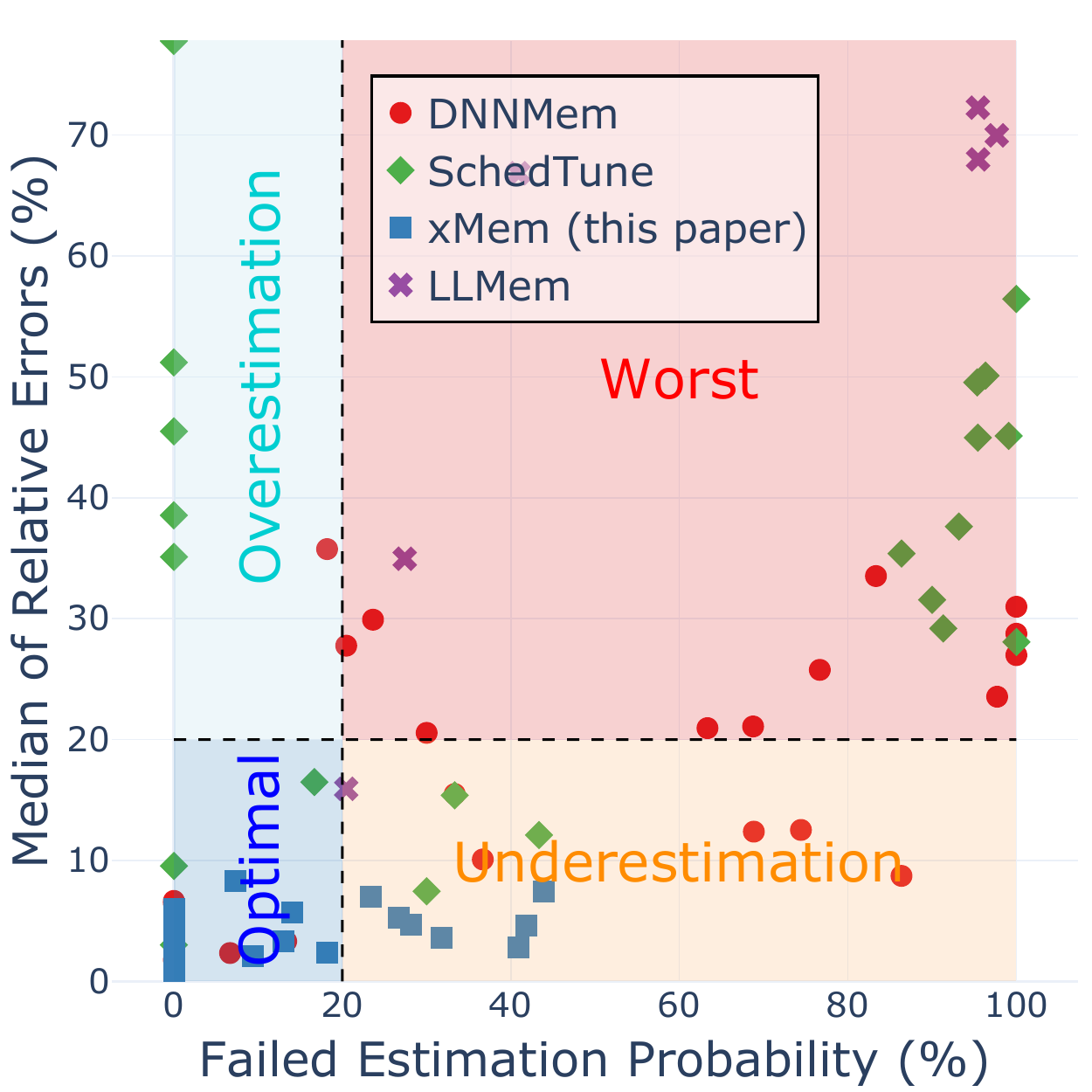}
    \caption{ANOVA Results}
    \Description{This scatter plot shows a four-quadrant analysis of Median Relative Error (MRE) versus Failed Estimation Probability (PEF) from detailed comparative runs (ANOVA-style)}
    \label{evaluation/fig/MREvPEF/ANOVA}
\end{subfigure}\hfill \begin{subfigure}{0.495\columnwidth}
    \includegraphics[width=\linewidth]{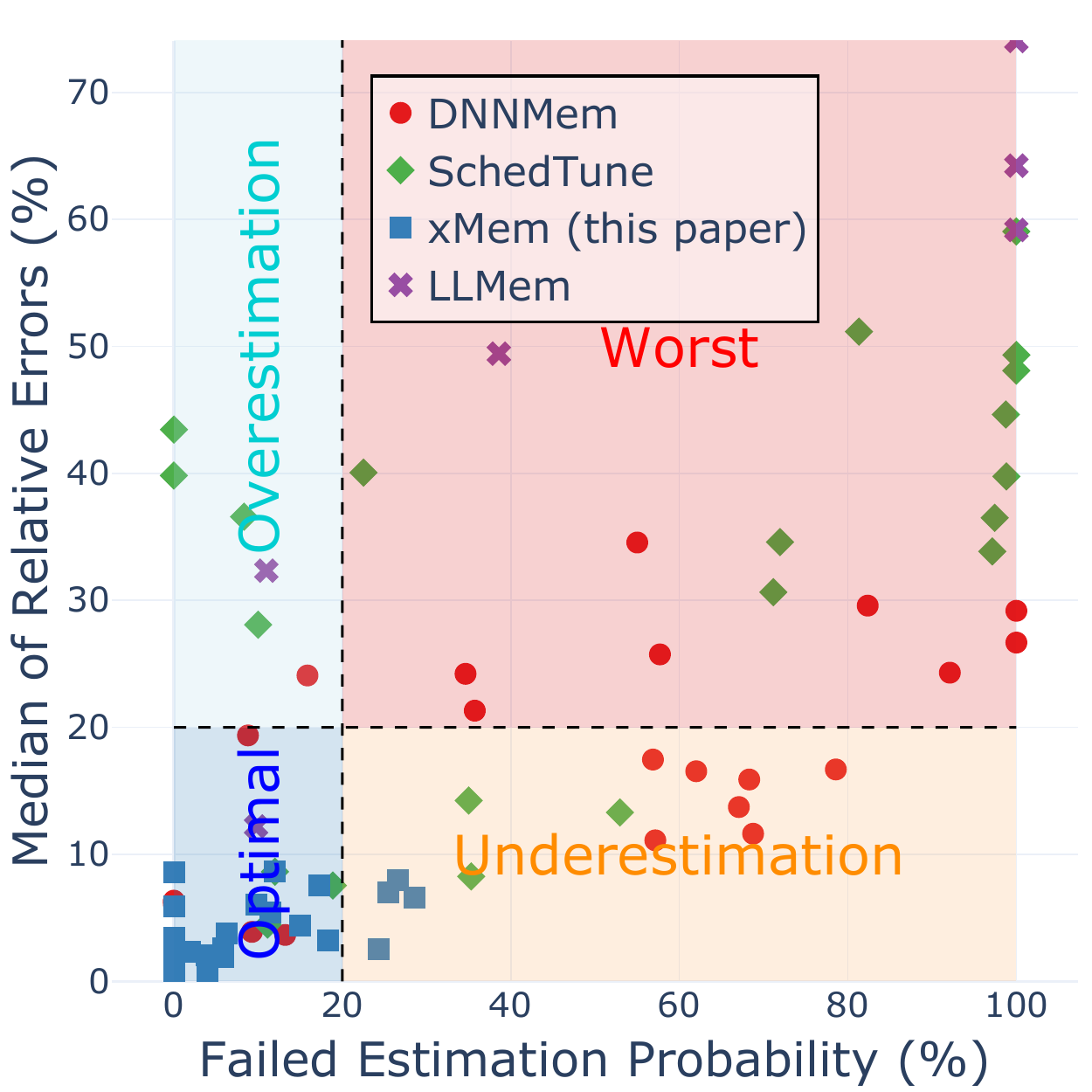}
    \caption{Monte Carlo Results}
    \Description{This scatter plot presents a four-quadrant analysis of Median Relative Error (MRE) versus Failed Estimation Probability (PEF) from large-scale Monte Carlo simulations}
    \label{evaluation/fig/MREvPEF/MC}
\end{subfigure}
\caption{Four-quadrant analysis of estimator performance, plotting the MRE against the PEF. Quadrants: Bottom left is {\color{fq-bl}optimal}; bottom right {\color{fq-br}underestimation}; top left {\color{fq-tl}overestimation}; and top right {\color{fq-tr}worst}.}
\label{evaluation/fig/MREvPEF/RQ2}
\Description{This scatter plot presents a four-quadrant analysis of Median Relative Error (MRE) versus Failed Estimation Probability (PEF) for both evaluation approaches: ANOVA and Monte Carlo}
\end{figure}

\subsection{RQ2: How reliable are the estimates of peak memory?}
\label{evaluation/chapter/rq/reliably}
Beyond accuracy, a crucial metric of a memory estimator is its reliability in practice; specifically, its ability to provide estimated memory that can be directly used as the maximum runnable memory without OOM, while not excessively overestimating. To answer RQ2, we analyze PEF, namely $P_{je2}$, primarily drawing from the aggregated results from both experiments. 
A lower PEF indicates higher reliability.

Figure~\ref{evaluation/fig/MREvPEF/ANOVA} presents a four-quadrant analysis plotting the PEF against the MRE for each estimator. The plot is divided by 20\% thresholds for both PEF (x-axis) and MRE (y-axis), creating four performance categories: Optimal (low PEF, low MRE), Overestimation (low PEF, high MRE), Underestimation (high PEF, low MRE), and Worst (high PEF, high MRE). As observed, \projectname's models (15 out of 22) dominantly cluster within the Optimal quadrant. Crucially, MRE for all models remains below 10\%, corroborating the results from RQ1. It is also remarkable that 12 of the 22 models satisfy criteria with both PEF and MRE values under 10\%. For the rest of the models presenting higher PEF, the persistently low MRE (<10\%) implies a high fidelity of memory estimation relative to ground truth values. This demonstrates a significant trade-off between accuracy and reliability. In contrast, DNNMem shows a considerable number of points scattered across the Underestimation and Worst quadrants, indicating a higher risk of OOM despite sometimes achieving acceptable MRE. SchedTune exhibits significant polarization; while two of its results are concentrated in the Optimal quadrant (although with MREs generally between 10\% and 20\%), a notable portion is clustered in the Worst quadrant, characterized by both high PEF and high MRE. LLMem shows a scattered distribution, with some points in less desirable quadrants.

Further validation through Monte Carlo simulations (detailed in Figure~\ref{evaluation/fig/MREvPEF/MC}) underscores the robustness of the proposed methods, showing enhanced relative performance beyond that observed in the four-quadrant ANOVA results. Specifically, the number of models in the Optimal quadrant increased from 15 to 18, with only MRE values below 10\%. Additionally, 13 models (up from 12) met the criteria with both PEF and MRE values under 10\%. In contrast, the baseline methods, by comparison, once more displayed significant instability or inherent performance shortcomings.

Therefore, in response to RQ2, \projectname demonstrates superior reliability in preventing OOM errors. Its ability to consistently provide estimates that are not only accurate but also reliable (low PEF) makes it a tool for practical GPU resource estimation compared to the evaluated baselines.

\subsection{RQ3: How much GPU memory can be conserved?} \label{evaluation/chapter/rq/memorysaving}

\begin{table}[htb]
\caption{Average MCP in Gigabytes (GB). Positive values indicate memory saved. `N/A' indicates non-applicability.}
\Description{This table shows the average memory saved in Gigabytes (GB) by different estimators (\projectname, DNNMem, SchedTune, LLMem) for Convolutional Neural Network (CNN) and Transformer model workloads, with `N/A' indicating non-applicability for an estimator-model combination.}
\begin{center}
\begin{tabular}{c|rrr|r}
\toprule
\textbf{Model Arch} & \textbf{DNNMem} & \textbf{SchedTune} & \textbf{LLMem} & \textbf{\projectname} \\
\midrule
CNN & 3.08 & 5.81 & N/A & 8.67 \\
Transformer & 1.29 & -4.42 & 1.68 & 7.07 \\
\midrule
Overall & 2.11 & 0.38 & 1.69 & 7.82 \\
\bottomrule
\end{tabular}
\label{evaluation/table/saved-memory}
\end{center}
\end{table}

Accurate and reliable memory estimation not only prevents OOM errors from wasting allocated GPU memory but also enables more efficient GPU memory utilization by avoiding unnecessary over-provisioning. To quantify this, we measure MCP as the average GPU memory saved per time, expressed as $\text{avgM}^{\text{save}}_{e}$. This metric incorporates a penalty for unreliable estimations that lead to OOM failures, as expressed by Eq.~\ref{evaluation/equation/memory-savings}. To ensure that MCP reflects real-world conditions across diverse and unpredictable scenarios, this section relies solely on the Monte Carlo simulation data for analysis.

Table~\ref{evaluation/table/saved-memory} summarizes average MCP for \projectname and the baseline estimators, categorized by CNN and Transformer model types. \projectname demonstrates substantial memory conservation across both categories. For CNN models, \projectname achieves an average MCP of 8.67 GB, significantly outperforming DNNMem (3.08 GB) and SchedTune (5.81 GB). LLMem was not applicable to CNNs as it only supports Transformer models. For Transformer models, \projectname conserves an average of 7.07 GB. This is considerably higher than DNNMem (1.29 GB) and LLMem (1.68 GB). Notably, SchedTune exhibits a negative mean MCP of -4.42 GB for Transformer models. This negative value signifies an average net loss of usable memory, attributable to OOM penalties resulting from its unreliability in this model category. Meanwhile, it implies a potential cold start problem of data-driven approaches, mentioned in Sections~\ref{introduction/chapter} and \ref{related/chapter/data-driven}.

The superior MCP achieved by \projectname is attributed to its high accuracy (low MRE) and outstanding reliability (low PEF), as established in answers to RQ1 and RQ2. By providing accurate estimates, \projectname allows tasks to be provisioned with memory resources much closer to their actual needs, thus freeing up substantial GPU memory that would otherwise be idly reserved or wasted due to inaccurate estimations. This translates directly to improved GPU utilization and the potential for higher task throughput in resource-constrained cluster environments.

\subsection{RQ4: What is the run-time overhead?} \label{evaluation/chapter/rq/runtime}

\begin{table}[htb]
\caption{The average runtime in seconds (s) for each estimator, including any input preprocessing.}
\Description{This table presents the average runtime for four methods: DNNMem (34.33 s), SchedTune (7.8 s), LLMem (18.24 s), and the proposed \projectname (24.73 s). SchedTune demonstrates the lowest runtime, indicating the highest operational efficiency.}
\begin{center}
\begin{tabular}{c|rrr|r}
\toprule
 & \textbf{DNNMem} & \textbf{SchedTune} & \textbf{LLMem} & \textbf{\projectname} \\
\midrule
\textbf{Runtime} & 33 & 2 & 17 & 26 \\
\bottomrule
\end{tabular}
\label{evaluation/table/runtime}
\end{center}
\end{table}

The run-time overhead for each memory estimation method was evaluated using Monte Carlo simulations. For data-analytical approaches such as \projectname and DNNMem, which rely on CPU-based analysis, this overhead was quantified as the CPU time taken to generate an estimate. In the case of SchedTune, the total overhead comprised the inference time of its pre-trained model plus the necessary input data preparation time. Finally, LLMem's overhead consisted of the time required for its targeted GPU execution measurements.

As detailed in Table~\ref{evaluation/table/runtime}, \projectname demonstrates an average estimation runtime of 26 seconds. While this performance is superior to that of DNNMem (33 seconds), a comparable data-analytical method, \projectname's runtime is longer than those of SchedTune (2 seconds), which leverages efficient pre-trained model inference, and LLMem (17 seconds), which performs direct GPU measurements. It should be noted that \projectname is currently in its prototype phase, and strategies for its optimization will be detailed in Section~\ref{discussion/chapter/runtime}. Despite the current speed, this runtime is deemed practical for pre-submission estimation, especially considering \projectname's substantial benefits in terms of accuracy, reliability, and memory conservation.

\subsection{RQ5: How accurately does the estimator estimate memory for large models?} \label{evaluation/chapter/rq/largermodel}

\begin{figure}[htbp]
    \centering
    \includegraphics[width=\columnwidth]{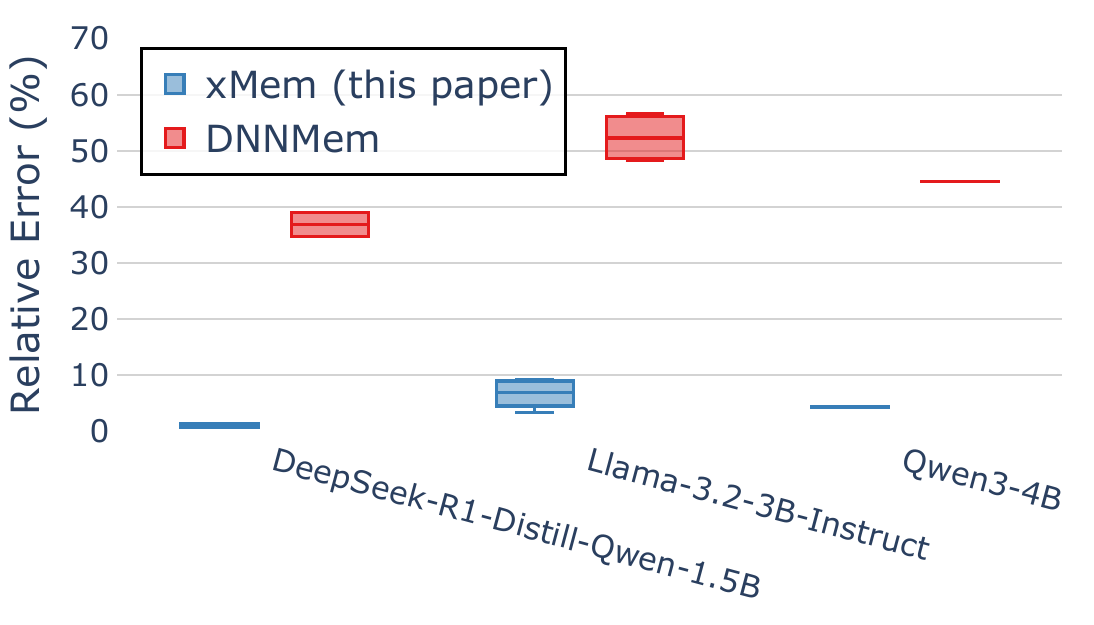}
    \caption{MRE comparison between \projectname and DNNMem for larger models on NVIDIA A100 GPU (Google CoLab).}
    \Description{This box plot chart shows the MRE percentage for \projectname's peak memory estimation on three large-scale models: DeepSeek-R1-Distill-Qwen-1.5B, Llama-3.2-3B-Instruct, and Qwen3-4B, when tested on an NVIDIA A100 GPU. \projectname (blue) shows low MRE for DeepSeek and Qwen3-4B, and a slightly higher MRE for Llama-3.2-3B-Instruct. DNNMem (red) is shown only for DeepSeek-R1-Distill-Qwen-1.5B, exhibiting a very high MRE.}
    \label{evaluation/figure/rq/largermodels}
\end{figure}

To assess \projectname's scalability and its accuracy with larger models, scenarios not feasible on our local GPUs due to memory limitations, we conducted evaluations using Google CoLab Pro instances equipped with NVIDIA A100 GPUs. Due to package conflicts preventing the reliable execution of SchedTune and LLMem in the CoLab environment, this analysis solely focuses on \projectname's performance compared to DNNmem. Additionally, due to the limited number of test configurations (explained in Section~\ref{eva/chapter/workload}) available for use in the CoLab experiments, it is not meaningful to present the PEF results. Therefore, we only show the MRE here.

Figure~\ref{evaluation/figure/rq/largermodels} illustrates the MRE of \projectname and DNNMem for these larger models on NVIDIA A100. \projectname demonstrates consistently strong accuracy and a significant advantage over DNNMem. For the Llama-3.2 (3B-Instruct) model, \projectname's MRE was 9.0\%, while a comparative figure for DNNMem was 52.3\%. On the DeepSeek-R1 (Distill-Qwen-1.5B) model, \projectname achieved an MRE of just 1.0\%, which is substantially lower than DNNMem's MRE of 37\%. Similarly, for the Qwen3 (4B) model, \projectname recorded an MRE of 4.3\%, again demonstrating superior performance compared to DNNMem's MRE of 44.6\%. These results underscore that \projectname's CPU-based dynamic analysis approach not only scales effectively to larger models but also performs accurately on advanced GPU architectures like the A100, consistently offering a clear advantage over DNNMem in terms of MRE. 
\section{Related Work}
\label{relatedwork/chapter}
The challenge of accurately estimating the peak GPU memory for DL jobs, particularly to prevent OOM errors and optimize resource utilization, has been introduced from several perspectives in the literature.

\subsection{Static Analysis Methods}
Static analysis methods derive memory estimates from the DL model's computational graph, offering a priori estimation without utilizing GPUs. DNNMem \cite{gao_estimating_2020}, a key baseline, implemented a detailed static approach by attempting to model the training process and simulating a framework-level memory allocator. Although this requires manual modeling for each operator's memory behavior, this is more comprehensive than simpler static methods like Horus \cite{yeung_horus_2022}, which primarily sums tensor sizes. However, the reliance of DNNMem on a static computation graph introduces several limitations. Such graphs inherently lack information regarding memory usage during the optimization phase, making estimations relatively more accurate for the lowest-overhead optimizers like SGD but less so for others. More critically, static graphs cannot capture dynamic memory variations resulting from code structure changes (\eg \texttt{optimizer.zero\_grad()} placement, Figure~\ref{motivation/fig/zero-out-memory-change}) or the precise runtime memory behavior of operators, leading to estimation inaccuracies as observed in our evaluation (Section~\ref{evaluation/chapter/rq/accurate}). Furthermore, while DNNMem simulates the framework's allocator, it does not model the device-level allocator or the crucial step of reclaiming cached segments that typically precedes an actual OOM event; true OOM occurs only when both allocator levels fail after reclamation attempts. \projectname, by contrast, employs dynamic analysis of CPU execution traces to obtain and model these runtime behaviors and allocator interactions directly.

\subsection{Data-Driven Estimation} \label{related/chapter/data-driven}
Another category of solutions uses ML models trained on historical data or model features. SchedTune \cite{albahar_schedtune_2022}, another baseline, and TBEM \cite{liu_tbem_2022} use pre-trained models to predict memory needs based on model and hardware characteristics. Although these methods can be fast at inference time, they typically require substantial and representative training datasets. More critically, they often face generalization challenges with new or significantly modified models (the cold start problem \cite{talha_deep_2023}) and may not precisely capture instantaneous peak memory demands or the effects of specific code-level configurations, thereby affecting their reliability (Sections~\ref{evaluation/chapter/rq/reliably} and \ref{evaluation/chapter/rq/memorysaving}). \projectname addresses this by directly analyzing the execution trace of the specific task at hand, rather than relying on generalized pre-trained models.

\subsection{Direct GPU Estimators}
Techniques performing estimation via direct GPU interaction, such as LLMem \cite{kim_llmem_2024} or Gandiva's \cite{222611} online profiling, may achieve high accuracy by measuring actual GPU usage. However, they face critical practical limitations: (i) GPUs' limited memory capacity restricts estimation for very large models that might not fit, a less severe issue for CPU-based methods as sufficient RAM can be provided in one server; (ii) direct GPU execution for estimation can itself trigger OOM errors; and (iii) the scarcity and high cost of GPU resources mean that using them for estimation incurs significant overhead, including resource contention and wait times. These drawbacks conflict with \projectname's goal of zero-GPU-overhead estimation, rendering such approaches less suitable for pre-submission checks for further scheduling decisions in GPU-constrained clusters.

\subsection{OOM Mitigation Techniques}
In contrast to a priori estimation, some approaches, such as AntMan \cite{258957} focus on runtime OOM mitigation by dynamically offloading tensors to the host RAM when the GPU memory is scarce. It is useful for GPU sharing and handling tasks exceeding the GPU capacity. However, these reactive methods can cause sustained performance degradation if a task's memory requirement exceeds the maximum of the GPU's capacity, forcing persistent offloading and frequent data transfers. \projectname's prior estimation is complementary, aiming to prevent OOM and such performance-degrading scenarios by enabling informed resource allocation.

\section{Discussion}
\label{discussion/chapter}
Our evaluation demonstrates that \projectname offers a significant advancement in prior GPU memory estimation, achieving high accuracy and reliability without target GPU overhead. In this section, we acknowledge several limitations and areas for future work. 

\subsection{Overhead} \label{discussion/chapter/runtime}
Regarding runtime overhead in Section~\ref{evaluation/chapter/rq/runtime}, \projectname's current CPU-based analysis (Table~\ref{evaluation/table/runtime}) involves processing massive profiling data, potentially containing millions of rows, and performing numerous time-range-based searches to link and attribute memory events. As a prototype, the current implementation processes this data sequentially and has not yet incorporated advanced big data processing techniques, such as a specific big-data processing library (e.g., Pandas, NumPy), concurrent data processing pipelines, or time-series databases for faster querying. These unoptimized search and indexing operations significantly contribute to the current runtime. While the memory simulation phase is comparatively less time-consuming, it also has potential for optimization. Future work will focus on these areas to reduce CPU overhead. Nevertheless, the crucial advantage of zero target-GPU usage during estimation remains, making its current runtime acceptable for pre-submission checks, especially given its superior accuracy and reliability.

\subsection{Distributed Training}

Although the current scope of \projectname is limited to single-GPU training, it forms the foundation for distributed systems. Practical distributed training relies on partitioning models and data, which requires precise memory data at either layer or operator level. While a large model cannot be fitted on a single GPU, we can leverage a CPU with abundant RAM to execute it and capture this essential per-layer/operator memory profile, a core concept of our method. This granular data forms the necessary foundation for any distributed estimation; without it, planning for model or pipeline parallelism would be based on guesswork. Moreover, the current architectural design (\S\ref{design/chapter/architecture}) is deliberately distribution-prepared (\S\ref{discuss/chapter/generalization}). A key future research direction is to investigate these distributed memory patterns and explore methodologies to extend \projectname, ideally by analyzing profile data obtained from a single node and then simulating distributed interactions.

\subsection{Mixed-precision Training}
Regarding mixed-precision training, since \projectname depends on profiling data, it is currently capable of estimating memory for FP16 precision when profiling data is available. Once an environment lacks supporting CPU instructions (\eg AVX2 VNNI 2 and AVX512 CORE FP16) \cite{intel_cpu_2024}, the profiling job itself can be time-consuming, especially for larger models. However, it is crucial to note that once the profiling data is collected, \projectname's subsequent analysis and estimation process remains the same. A future direction utilizes profiling data, providing key properties such as input and output tensor dimensions. During FP32 or FP16 training, the tensor dimensions remain constant; only the data type changes. Hence, the analyzer can deduce memory usage by identifying and filtering memory blocks affected by input tensors. This approach could quickly adopt a new type of data, such as FP8.

\subsection{Generalization and Scalability} \label{discuss/chapter/generalization}

Given PyTorch’s prevalence in both industry \cite{ryan_pytorch_2023, ana_pytorch_2025} and research \cite{ryan_pytorch_2023, daniel_tensorflow_2025}, \projectname's current version can serve the majority of cluster workloads. However, to ensure future extension, the framework was architecturally designed with extensibility for multiple DL frameworks and hardware devices, as reflected in its distinct Analyzer, Orchestrator, and Simulator components. This design makes it a versatile solution.
\begin{enumerate*}[label=(\roman*)]
    \item \textbf{Distribution-Prepared}: The Memory Orchestrator can leverage per-layer data from the Analyzer to model pipeline schedules or inject simulated allreduce buffers, thereby preparing for future distributed training scenarios.
    \item \textbf{General BFC Algorithm}: Our core simulation is based on the BFC algorithm, which is not exclusive to PyTorch and is also used by frameworks like TensorFlow \cite{tensorflow_tensorflow_2025, nvidia_corporation_tensorflow_2025} for CUDA memory management.
    \item \textbf{Pluggable Architecture}: The Analyzer, Orchestrator, and Simulator components are designed to be framework- and hardware-agnostic, allowing for new modules to be implemented for different profiling outputs, device behaviors, and memory allocation algorithms. Notably, accurately modeling each allocator is crucial for accurate memory estimation, as generic models often overlook nuanced behaviors. For example, a caching allocator might request a 20MB block for a 10MB tensor (\S\ref{background/chapter/memory}), with an Out-of-Memory (OOM) error occurring only after failing to reclaim cached memory.
\end{enumerate*} 

\section{Conclusion}
\label{conclusion/chapter}
This paper addresses the challenge of resource mismatch in shared GPU clusters, which frequently results in OOM errors and resource underutilization, thereby wasting costly GPU resources. To mitigate this, we leverage a key characteristic of modern DL frameworks like PyTorch: their architectural decoupling of front-end and back-end components. Specifically, the high-level logic defined in a Python training script determines a consistent number of core tensors and a fixed execution sequence, irrespective of the underlying hardware device (\eg CPU or GPU). Upon this, we propose \projectname, a novel estimation framework. It focuses on orchestrating the differences in memory behavior between CPU and GPU operators by analyzing and adjusting memory event timings in CPUs. This enables an accurate and reliable estimation of GPU memory requirements based solely on CPU profiling data.

Extensive experimentation demonstrates \projectname's outstanding performance over SOTA solutions, significantly improving MRE by \xmemmedianerrorimprove and reducing the PEF by \xmemprobabilityimprove. This accurate and reliable estimation enables cluster schedulers incorporating \projectname to realize substantial memory savings for the GPU cluster, evidenced by a \xmemmemoryimprove increase in MCP compared to existing solutions. By allowing for more precise resource scheduling, these savings directly contribute to mitigating the persistent issue of GPU scarcity in cloud providers, research institutions, and enterprises.

\balance
\bibliographystyle{ACM-Reference-Format}

\end{document}